\documentstyle[12pt,epsfig,axodraw]{article}
\advance\textheight by \topskip
\begin{document}
\tolerance=100000
\thispagestyle{empty}
\setcounter{page}{0}

\newcommand{\be}{\begin{equation}}
\newcommand{\ee}{\end{equation}}
\newcommand{\br}{\begin{eqnarray}}
\newcommand{\er}{\end{eqnarray}}
\newcommand{\ba}{\begin{array}}
\newcommand{\ea}{\end{array}}
\newcommand{\bi}{\begin{itemize}}
\newcommand{\ei}{\end{itemize}}
\newcommand{\bn}{\begin{enumerate}}
\newcommand{\en}{\end{enumerate}}
\newcommand{\bc}{\begin{center}}
\newcommand{\ec}{\end{center}}
\newcommand{\ul}{\underline}
\newcommand{\ol}{\overline}
\newcommand{\ar}{\rightarrow}
\newcommand{\sm}{${\cal {SM}}$}
\newcommand{\as}{\alpha_s}
\newcommand{\aem}{\alpha_{em}}
\newcommand{\ycut}{y_{\mathrm{cut}}}
\newcommand{\susy}{{{SUSY}}}
\newcommand{\Dir}{\kern -6.4pt\Big{/}}
\newcommand{\Dirin}{\kern -10.4pt\Big{/}\kern 4.4pt}
\newcommand{\DDir}{\kern -10.6pt\Big{/}}
\newcommand{\DGir}{\kern -6.0pt\Big{/}}
\def\Ecm{\ifmmode{E_{\mathrm{cm}}}\else{$E_{\mathrm{cm}}$}\fi}
\def\gluino{\ifmmode{\mathaccent"7E g}\else{$\mathaccent"7E g$}\fi}
\def\photino{\ifmmode{\mathaccent"7E \gamma}\else{$\mathaccent"7E \gamma$}\fi}
\def\mgluino{\ifmmode{m_{\mathaccent"7E g}}
             \else{$m_{\mathaccent"7E g}$}\fi}
\def\taugluino{\ifmmode{\tau_{\mathaccent"7E g}}
             \else{$\tau_{\mathaccent"7E g}$}\fi}
\def\mphotino{\ifmmode{m_{\mathaccent"7E \gamma}}
             \else{$m_{\mathaccent"7E \gamma}$}\fi}
\def\ML{\ifmmode{{\mathaccent"7E M}_L}
             \else{${\mathaccent"7E M}_L$}\fi}
\def\MR{\ifmmode{{\mathaccent"7E M}_R}
             \else{${\mathaccent"7E M}_R$}\fi}

\def\Ord{\buildrel{\scriptscriptstyle <}\over{\scriptscriptstyle\sim}}
\def\OOrd{\buildrel{\scriptscriptstyle >}\over{\scriptscriptstyle\sim}}
\def\pl #1 #2 #3 {{\it Phys.~Lett.} {\bf#1} (#2) #3}
\def\np #1 #2 #3 {{\it Nucl.~Phys.} {\bf#1} (#2) #3}
\def\zp #1 #2 #3 {{\it Z.~Phys.} {\bf#1} (#2) #3}
\def\pr #1 #2 #3 {{\it Phys.~Rev.} {\bf#1} (#2) #3}
\def\prep #1 #2 #3 {{\it Phys.~Rep.} {\bf#1} (#2) #3}
\def\prl #1 #2 #3 {{\it Phys.~Rev.~Lett.} {\bf#1} (#2) #3}
\def\mpl #1 #2 #3 {{\it Mod.~Phys.~Lett.} {\bf#1} (#2) #3}
\def\rmp #1 #2 #3 {{\it Rev. Mod. Phys.} {\bf#1} (#2) #3}
\def\sjnp #1 #2 #3 {{\it Sov. J. Nucl. Phys.} {\bf#1} (#2) #3}
\def\cpc #1 #2 #3 {{\it Comp. Phys. Comm.} {\bf#1} (#2) #3}
\def\xx #1 #2 #3 {{\bf#1}, (#2) #3}
\def\preprint{{\it preprint}}

\begin{flushright}
{RAL-TR-1999-017}\\ 
{December  1999}\\
\end{flushright}

\vspace*{\fill}

\begin{center}
{\Large \bf 
Six-quark decays of off-shell $W^+W^-$ pairs \\[0.25 cm]
via parton-level QCD\\[0.5 cm]
in electron-positron annihilations}\\[0.5 cm]
{\large S. Moretti$^*$}\\[0.4 cm]
{\it Rutherford Appleton Laboratory,}\\
{\it Chilton, Didcot, Oxon OX11 0QX, UK.}\\[0.5cm]
\end{center}
\vspace*{\fill}

\begin{abstract}
{\noindent 
We study the decays  into six quarks
of off-shell pairs of $W^\pm$ bosons produced in 
electron-positron 
annihilations, through the ${\cal O}(\alpha_s^2)$ order
in the strong coupling constant. We give explicit helicity amplitude
formulae. We present numerical results in the context of phenomenological 
analyses of relevance at LEP2 and future
Linear Colliders: such as $M_W$ determinations, 
New Physics
and Higgs searches.}
\end{abstract}
\vskip10.0cm
\hrule
\vskip0.25cm
\noindent
$^*$ E-mail: moretti@rl.ac.uk.
\vspace*{\fill}
\newpage

\section{Introduction}
\label{sec_intro}

Pairs of $W^\pm$ bosons have  copiously been produced at LEP2
and studied in great detail by the four experimental Collaborations
over the past few years. In fact, 
one of the main goals of such collider is the determination of $M_W$ with a 
target accuracy of 50 MeV or less.
One of the detection strategies adopted to measure the $W^\pm$ mass
is the kinematic reconstruction of the $W^\pm$ resonances through the
momenta of their decay products, in the fully hadronic channel: 
$W^+W^-\ar$ jets.
Although the event reconstruction is made hard in this decay mode
by the large number of tracks in the detector and by the usual 
uncertainties related to measuring jet energies and directions, a task 
much less complicated in the case of semileptonic decays, 
$W^+W^-\ar$ 2 jets $\ell^\pm$ plus
missing energy (with $\ell=e,\mu,\tau$), and despite the existence 
of theoretical biases  due to the
relatively unknown `Bose-Einstein correlations' \cite{bose} and 
`colour-rearrangement' \cite{CR} effects, from  
which the semi-leptonic decays are immune, hadronic decays of
$W^+W^-$ pairs have been successfully exploited at LEP2. On the one hand, 
they yield the largest decay rate. On the other hand, 
the knowledge of all momenta in the final state helps tightening 
the $W^\pm$ mass resolution.

The problem with the fully hadronic mode is twofold. 
Firstly, because two identical decays take place in the same event, one has
the phenomenon of mis-pairing of jets. That is,  even
in the ideal case in which all hadronic tracks are correctly assigned
to the parton from which they originate, one has to cope with the ambiguity
that it is in practise impossible to uniquely assign any pair among the four 
reconstructed jets to the parent $W^\pm$ on the sole basis of the 
event topology.  Of all possible combinations of di-jet systems, only one is 
correct. Thus, an
intrinsic background exists in $W^+W^-\ar 4$ jet events, in terms of simple
combinatorics. Secondly, because of the large hadronic multiplicity,
one also has the phenomenon of mis-assignment of tracks. This is induced by
the procedure adopted in selecting jets. This is generally done by resorting 
to so-called jet
clustering algorithms \cite{schemes}, wherein the number 
of tracks is reduced one at a time by combining the two most (in some sense)
nearby ones. The joining procedure is stopped by means of a resolution 
parameter,  $\ycut$, and the final `clusters' are called jets. Here,
the ambiguity  stems from the fact that a track assigned to a cluster, 
the latter eventually identified as the  parton
originating from one of the $W^\pm$'s, might have actually been produced 
in the fragmentation of another parton coming from the second $W^\pm$ decay.

In both cases, the consequence is a distortion of the `line shape' of the
$W^\pm$ resonance, which needs to be accurately quantified 
if one wants to achieve the foreseen precision in the $W^\pm$ mass 
measurement. 
In order to estimate these effects, one can resort to  
phenomenological Monte Carlo (MC) programs (e.g.,
HERWIG \cite{HERWIG}, JETSET/PYTHIA \cite{JETSET}
and ARIADNE \cite{ARIADNE}). These represent a valuable instrument 
in this respect, as they are able to 
describe the full event, from the initial hard scattering down to the hadron 
level. On the other hand, Matrix Element (ME) models are acknowledged to 
describe the large angle distributions of the QCD (and QED) radiation
better than the former (see, e.g., \cite{ALEPHgluino,spin}), which are in 
fact superior in the small angle dynamics. 
Whereas the above MC programs have been in use for long time, so that their
features  need not being recalled 
 here, it might be useful to review at this stage
the progress made in ME calculations of $e^+e^-\ar W^+ W^-\ar$ jets,  as many 
of the achievements in this field are very recent.

For a start, it should be mentioned that 
the amplitude for $e^+e^-\ar W^+W^-\ar q\bar q q'\bar q'$ is 
very trivial to derive, in fact, more of a textbook example
(here and in the following,
$q$, $q'$ and $q''$ refer to (anti)quarks produced in the $W^\pm$, $W^\mp$
and $g$ splittings, respectively). It represents 
the lowest-order (LO) 
contribution to the $W^+W^-\ar 4$ $\mbox{jet}$ hadronic signal. Higher-order 
(HO) effects are those involving gluon emission: for example, the real one 
(i.e., a tree-level contribution) in $e^+e^-\ar W^+W^-\ar q\bar q q'\bar q' g$
and $e^+e^-\ar W^+W^-\ar q\bar q q'\bar q' gg$ events, which have been 
calculated in Refs.~\cite{Brown} and \cite{Torino}, respectively.
One-loop QCD corrections to $e^+e^-\ar W^+W^-\ar q\bar q q'\bar q'$
are also known to date \cite{loop}, and they have been combined with the LO 
emission of Ref.~\cite{Brown} into the complete ${\cal O}(\alpha_s)$ result
\cite{loop}. 

It is purpose of this paper to study  the reaction  
$e^+e^-\ar W^+W^-\ar q\bar qq'\bar q'q''\bar q''$ at tree-level, 
through the order ${\cal O}(\alpha_{em}^4\alpha_s^2)$.
Earlier accounts of this process, with emphasis on it being a
background to top-antitop production and decay at a future
Linear Collider (LC), were given in Ref.~\cite{WW6q}. 
The complete set of Feynman diagrams
needed to perform such a calculation can be found in Fig.~\ref{fig_graphs}.
The plan of the paper is as follows. In the next Section we 
highlight the phenomenological impact at LEP2 and a LC
of six-quark production via $W^+W^-$ decays. 
In Sect.~\ref{sec_calc}  
we describe our method of computation using tree-level 
perturbative QCD. In 
Sect.~\ref{sec_results} 
we present our results whereas Sect.~\ref{sec_summary} 
contains our conclusions.

\section{Phenomenology of six-quark decays of $W^+W^-$ pairs}
\label{sec_motiv}

In our opinion, there are a few good reasons to tackle the calculation
of the process $e^+e^-\ar W^+W^-\ar q\bar qq'\bar q'q''\bar q''$.

As for $M_W$ determinations, one should note the following. 
\begin{enumerate}
\item 
The availability of all HO corrections
to the leading four-parton decay of $W^\pm$ pairs is essential,
both at tree and loop level, for the following reasons.
On the one hand, from the point of view of perturbative calculations, 
it is evident that the 4-jet rate is constituted not only of 
the fraction  of 4-parton
events in which  all partons are resolved (i.e., their separation is above 
the cut-off $\ycut$), but 
also by the $(4+m)$ parton configurations (with $m\ge1$) in which
$m$ partons remain unresolved. 
On the other hand, because of the way an experimental 4-jet sample of $W^+W^-$
events is normally selected (see, e.g., \cite{ALEPH}), also the  
$(4+m)$ parton contributions with $m$ jets resolved 
 are relevant, as they naturally enter the candidate experimental sample
of hadronic decays of $W^+W^-$ pairs.
In fact, in order to maximise the event rate of the signal, one 
usually requires to reconstruct {\sl at least} four jets, all with separation 
above a minimum $\ycut$. Only 
eventually these jets are forced into exactly four, by merging together 
 those which are `closer'. In this respect, one subtlety should be noted
concerning $W^+W^-\to 6$~quark events.
The diagrams shown in Fig.~\ref{fig_graphs} implicitly assume
that the flavour of the quarks produced in the gluon splitting, labelled
by 7 and 8, is different from that of any of the fermions generated by 
the $W^\pm$ decays, indicated by 3, 4, 5 and 6. In fact, in case one or two 
quarks (or antiquarks) are identical, the numbers of diagrams doubles or 
quadruples, respectively, the new graphs being obtained from the old ones
by exchanging one or two identical fermion legs, in all possible ways
(a minus sign factorises too, for any of these exchanges). This 
follows from the fact that the flavour of a quark and its origin
(i.e., whether it comes from a $W^\pm$ or a $g$) are indistinguishable in the
experimental hadronic sample.
From the point of view of $W^\pm$ mass studies, it is evident that this is
source of two types of systematic effects. Firstly,
the total rate of $W^+W^-\ar6$~quark decays naively obtained by summing 
over five flavours in the $g^*\ar q''\bar q''$ splitting (the approach
used in parton shower models) could be mis-estimated. Secondly,
additional distortion effects to the line-shape of the 
$W^\pm$ resonance could occur.
In the spirit of Refs.~\cite{bose,CR}, 
the reader should not object to calling such effects  
`flavour-rearrangement' or `Fermi-Dirac correlations'.
\vskip0.25cm\hskip-0.25cm
\noindent
\hskip-0.25cm{We further focus on possible 
New Physics and Higgs boson searches}.
\item The very fact that one of the two $W^\pm$'s has a chance to decay
into four jets reproduces at LEP2 and a LC
a dynamics similar to that analysed
in several instances at LEP1, when studies of $\gamma^*,Z\ar4$ jet decays
were performed\footnote{Not quite, it could well be argued, given that 
in $W^+W^-\to6$~quark decays one
has to first isolate a subset of four jets which come from the same 
$W^\pm$ decay, out of the original six, and since such an operation is in 
principle affected by the same mis-assignment problems already described.
In practise, we will show that is rather easy to select such a subject,
the latter preserving the typical angular properties of gauge 
bosons and fermions
exploited in the experimental fits (see discussion later on).}, 
in view of the possible existence of New Physics contributions
due light gluinos ${\mathaccent"7E g}$ of the MSSM
\cite{window}. 
The evidence of such effects at LEP1  
was mainly
searched for in the context of the measurement of the three fundamental
colour factors of QCD: $C_A$, $C_F$ (the Casimir operators of the fundamental
and adjoint  representations of
the gauge group $SU(N_C)$, respectively) and $T_F$ (the normalisation of the
generators of the fundamental representation). 
In fact, under the assumption that $SU(N_C\equiv3)$ is indeed the
gauge group of QCD, with $N_C$ the number of colours, a measurement of these 
parameters (in particular of $T_R=N_FT_F$)
can be converted into a constraint on 
the number of coloured fermions active at the energy scale at which the 
decay  takes place. For example, $N_F$ would be
increased (by approximately 3) from its 
SM value at LEP energies (i.e., $N_F=5$) by the additional presence of
gluinos produced via a $g\rightarrow
{\mathaccent"7E g}{\mathaccent"7E g}$ splitting \cite{epem}.
In ordinary QCD, one gets $C_A=3$ and $C_F=4/3$ (also,  $T_F=1/2$).
Since the results of these analyses have not excluded the possible 
existence of Supersymmetric (SUSY) events in LEP1 data
in the form of {\sl very} light gluinos, with mass below 1--1.5 
GeV \cite{ALEPHgluino,Glennys1}, 
we consider whether contributions of
$e^+e^-\ar W^+W^-\ar q\bar q q'\bar q' {\mathaccent"7E g}{\mathaccent"7E g}$
events
possibly entering the six-jet sample produced at LEP2 and a LC can at all 
be disentangled. (Notice that
the above mass region has escaped also the LEP1 limits imposed through
the running of $\alpha_s$, at the three-loop 
perturbative level \cite{Csi}, in hadronic decays of heavy particles
\cite{observables} (e.g., $Z$ and $\tau$), as
well as those obtained from other experiments (KTeV, NA48, E761) searching
 for decays of gluino bound states\footnote{Some possible, Supergravity
(SUGRA)
inspired decay modes are \cite{gluonia}: 
$R^0\ar \pi^+ \pi^-{\tilde\gamma}$ and $R_p\ar S^0\pi^+$,
where $R^0$ is the so-called glueballino bound state 
$R^0\equiv(g\gluino)$ and the `photino' $\photino$ is the lightest SUSY 
particle (LSP), whereas $R_p\equiv(uud\gluino)$ and $S^0\equiv(uds\gluino)$,
with $S^0\to R^0\Lambda$. 
The mass(lifetime) of the $R^0$
is set by the theory to be in the range $1.4-2.2$ GeV($10^{-5}-10^{-10}$ sec) 
whereas that of the $R_p$ 
is $1.6-3.1$ GeV($2\cdot10^{-10}-2\cdot10^{-11}$ sec).
For alternative decay modes, in the Gauge Mediated SUSY 
breaking scenario, which have
recently been proposed and not yet exploited experimentally, 
see Ref.~\cite{newgluonia}.}, 
see \cite{Glennys_review}.)
The much reduced $W^+W^-$ cross section at LEP2 and a LC,
as compared to the $Z$
one at LEP1, clearly disfavours such a possibility. 
In contrast, it would be intriguing to consider a 
selection strategy similar to that advocated in Ref.~\cite{noigluino}, 
based on the detection of a secondary vertex possibly due to a  gluino decay
 with lifetime between, 
say, $10^{-12}$ (the typical scale of $b$  and $c$ quarks) and $10^{-9}$ (the 
coverage of the LEP detectors) sec. Besides,  the fact that 
a $W^\pm$ cannot decay directly into $b$ quarks (apart
from Cabibbo-Kobayashi-Maskawa suppressed channels, that we 
neglect here) implies that 
at LEP2 and a LC the overwhelming background from ordinary
QCD due to $e^+e^-\ar W^+W^-\ar q\bar q q'\bar q' gg$ events,
with $q^{(')}\ne b$, should effectively be 
removed by asking for just a single vertex tagging
 displaying
a decay lenght comparable or longer than that induced by $b$ quarks,
 $d_b\approx0.3$ mm.
This way, the surviving six-jet sample would only be composed by  
          $e^+e^-\ar W^+W^-\ar q\bar q q'\bar q' b\bar b$ and
          $e^+e^-\ar W^+W^-\ar q\bar q q'\bar q' \gluino\gluino$ 
events. These have comparable production rates, which 
is a most welcome result. 
After all, the smallness of the SUSY effect with respect to ordinary
QCD dynamics was really the limiting 
factor of the experimental analyses performed at LEP1, 
given that even the very large next-to-leading-order
(NLO) corrections to the four-jet sample 
leave the shape of the distributions used to fit $C_A$, $C_F$ and $T_F$
practically unaltered \cite{anglesNLO}.  For all the above reasons,
we believe it then important looking into this aspect of six-jet
 phenomenology. 

\item Finally, we
consider the possibility that six-quark decays of $W^+W^-$ pairs
with a gluon splitting into $b\bar b$ pairs, i.e., 
$e^+e^-\ar W^+W^-\ar q\bar q q'\bar q' b\bar b$ events, 
can act as a background to Higgs signals possibly produced  at a LC. 
In the Standard Model (SM) one has the production 
mechanism $e^+e^-\ar \phi Z$, whereas in the Minimal Supersymmetric Standard
Model (MSSM) one can have two,
$e^+e^-\ar \Phi Z$ and $e^+e^-\ar \Phi A$, where $\Phi=H,h$ represents
any of the two scalar Higgs bosons of the SUSY theory and $A$ is the 
pseudoscalar one. If one recalls that $Z$ bosons decay into bottom quark pairs
some 15\% of the times and that the SM Higgs decay channel $\phi\ar W^+W^-$ 
is the one with largest branching ratio (BR) for $M_\phi\OOrd140$ GeV, 
then it is evident that a very large
part of the SM Higgs signal at a LC would appear through
a six-jet signature, further considering that each $W^\pm$ boson
decays hadronically with a 70\% BR. Similarly, in the 
MSSM\footnote{Assuming that the mass scale of SUSY
partners of ordinary matter is in the TeV range, as preferred
by many Grand Unification Theories (GUTs).},
the $H\ar W^+W^-$ (of the heavy scalar Higgs) decay rate can be very large
on a big portion of the $(M_A,\tan\beta)$ plane (especially for
Higgs masses below the top-antitop decay threshold and 
at low $\tan\beta$ \cite{ioejames}) and so is the case of the 
$h\ar W^+W^-$ (of the light scalar Higgs) one, though for a more 
restricted area of the MSSM parameter space. Furthermore, $A\ar b\bar b$
decays of the MSSM pseudoscalar Higgs are always dominant, except
at low $\tan\beta$  and large $M_A$ (above 200 GeV). 
Indeed, considering that in the fully hadronic channel,
in order to suppress both the effects of combinatorics and of reducible  
backgrounds (e.g., 
$e^+e^-\ar \gamma^*,Z^*\ar q\bar qgggg$ \cite{ee6j}), vertex tagging
techniques 
will certainly be exploited to select $b\bar b$ pairs produced 
in $Z$ and $A$ decays, then it is quite likely that six-quark decays of 
$W^\pm$ pairs
(with two $b$'s) can be a serious
noise. Even more so if one further realises that these are irreducible in two
respects: not only in the particle content of the final state, but also
because they naturally contain a di-jet pair resonating at the $W^\pm$ mass.
\end{enumerate}

\section{Calculation}
\label{sec_calc}

\begin{figure}[p]
~\epsfig{file=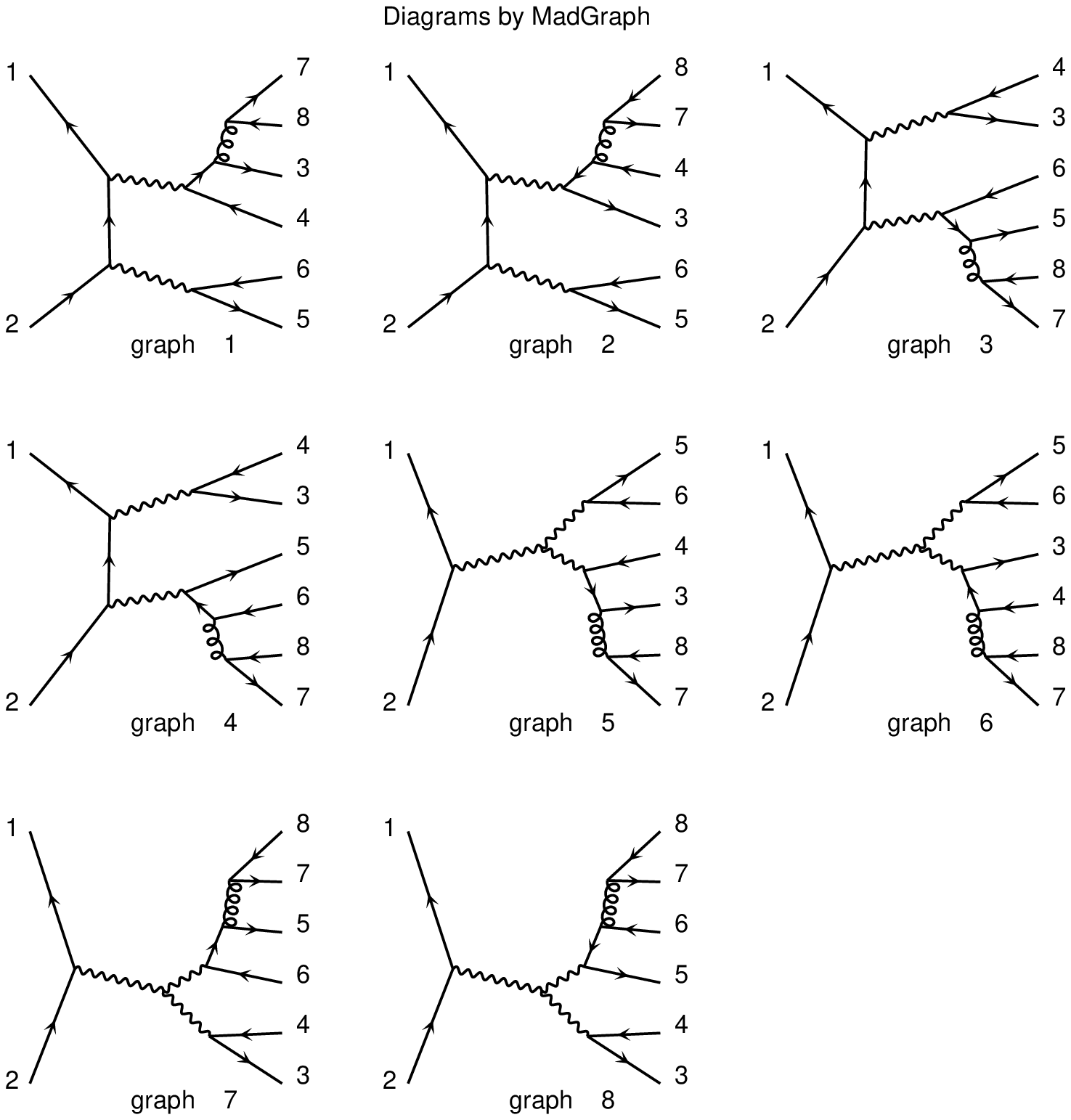,height=20cm}
\vskip-6.0cm
\caption{Feynman diagrams contributing to
$e^+(1) + e^-(2)$ $\ar$ $q(3)$ + $\bar q(4)$ + 
                                 $q'(5)$ +  $\bar q'(6)$ +
                                $q''(7)$ + $\bar q''(8)$,
via $W^+W^-$ production and decay, through
the order ${\cal O}(\aem^4\as^2)$. 
Here, we assume $q''\ne q',q$. An internal wavy
line represents a $\gamma,Z$ or a $W^\pm$, as appropriate.}
\label{fig_graphs}
\end{figure}

In the numerical part of our calculations, 
as centre-of-mass (CM) energies representative of LEP2 and a future LC, we have
used the values $\Ecm=172$ and 350, 500 GeV, respectively. As for the 
parameters of the theory,
we have adopted $M_Z=91.17$ GeV, $\Gamma_Z=2.516$ GeV,
                $M_W=80.23$ GeV, $\Gamma_W=2.2$ GeV, 
$\sin^2 \theta_W=0.23$, $\alpha_{{em}}= 1/128$ and the two-loop expression
for $\alpha_{s}$, with 
$\Lambda_{\tiny{\mbox{QCD}}}^{N_F=5}=0.200$ GeV. Furthermore, 
we have kept all quarks massless as a default except the bottom ones
(for which we used $m_b=4.95$ GeV), in order to speed up the numerical 
evaluations. 
(Electron and positron have mass zero too, so have the neutrinos.)
In practise, as we neglect Cabibbo-Kobayashi-Maskawa mixing terms
(see below, Tab.~\ref{tab_cRcL}),
this corresponds to always neglect the mass along the fermion lines 
connected to 
the $W^\pm$ currents in Fig.~\ref{fig_graphs}, while keeping the one of those
emerging from the gluon. This should be a good approximation.
On the one hand, the energy produced in the $W^\pm$ decays both at LEP2 and
a LC is typically much larger than the quark masses. On the other hand,
the latter can give sizable effects in the splitting of the
gluon \cite{noimass}. This approach also naturally
allows us to study effects of massive gluinos. As for these,
we have spanned their mass $m_\gluino$ 
over the range 0 to 10 GeV. While doing so, 
we also have to compare the six-quark and four-quark-two-gluino rates with 
the four-quark-two-gluon ones. That is, we have to calculate
the process $e^+e^-\ar W^+W^-\ar q\bar q q'\bar q' gg$. To do so,
we have 
resorted to the {\tt HELAS} \cite{HELAS} subroutines, whose results agree with
those reported in Ref.~\cite{Torino}. 

In addition, 
when $e^+e^-\ar {\mbox{Higgs}}~Z\ar W^{+(*)}W^{-(*)} b\bar b \ar bb~+4$~jets 
and  $e^+e^-\ar {\Phi}~A        \ar W^{+(*)}W^{-(*)} b\bar b \ar bb~+4$~jets 
have been  calculated (at tree-level), where ${\mbox{Higgs}}=\phi,\Phi$,
with $\phi$  the SM
scalar boson and $\Phi=H$, $h$ the corresponding ones in the MSSM,
we have used again heavy $b$ quarks, with their mass appropriately run
up to the mass of the decaying pseudoscalar $A$ boson
in the second process,
to match the procedure employed to compute the total widths of the latter
\cite{ioejames}. Mass relations and couplings involving the 
neutral MSSM Higgs bosons
have been computed using the full one-loop and the leading two-loop
corrections \cite{carena}, using $\tan\beta$ and $M_A$ as inputs,
 assuming  a universal soft Supersymmetry-breaking mass of 1 TeV  
and negligible mixing in the stop and sbottom mass matrices.
To calculate the Higgs cross sections and differential rates we have
used the exact $2\ar6$ MEs, including finite width effects of all 
unstable particles, that we have produced using 
again the {\tt HELAS} subroutines.

In order to calculate the Feynman diagrams in Fig.~\ref{fig_graphs}
we have used {\sl two} different spinor methods, this enabling us to check
the correctness of our results.
The first one is based on the formalism of Ref.
 \cite{HZ}. 
A second approach is based on the method developed in    
Refs.~\cite{KS,mana,ioPRD}.  
Since we will express the helicity amplitudes in this last formalism, we
will devote same space here to describe its technicalities. 

\begin{table}[!t]
\begin{center}
\begin{tabular}{|c|c|}
\hline
$\lambda_1\lambda_3$ &
$X(p_1,\lambda_1;p_2;p_3,\lambda_3;c_R,c_L)$ \\ \hline\hline
$++$  & $(\mu_1\eta_2 + \mu_2\eta_1)
(c_R\mu_2\eta_3 + c_L\mu_3\eta_2) + c_R S(+,p_1,p_2) S(-,p_2,p_3)$ \\ \hline
$+-$ & $c_L(\mu_1\eta_2 + \mu_2\eta_1)   S(+,p_2,p_3)
      + c_L (c_L\mu_2\eta_3 + c_R\mu_3\eta_2 )S(+,p_1,p_2)$ \\ \hline
\end{tabular}
\caption{The $X$ functions for the two independent helicity combinations
in terms of the functions $S$, $\eta$ and $\mu$ defined in the text.
The remaining $X$ functions can be obtained by flipping the sign
of the helicities
and exchanging $+$ with $-$ in the $S$ functions and $R$ with $L$ in the
chiral coefficients.}
\label{tab_X}
\end{center}
\end{table}

\begin{table}[!ht]
\begin{center}
\begin{tabular}{|c|c|}
\hline
$\lambda_1\lambda_2\lambda_3\lambda_4$ &
$Z(p_1,\lambda_1;p_2,\lambda_2;p_3,\lambda_3;p_4,
\lambda_4;c_R,c_L;c'_R,c'_L)$\\\hline \hline
$++++$ & $-2[S(+,p_3,p_1)S(-,p_4,p_2)c'_Rc_R
            -\mu_1\mu_2\eta_3\eta_4c'_Rc_L
            -\eta_1\eta_2\mu_3\mu_4c'_Lc_R]$  \\ \hline
$+++-$ & $-2\eta_2c_R[S(+,p_4,p_1)\mu_3c'_L-S(+,p_3,p_1)\mu_4c'_R]$ \\ \hline
$++-+$ & $-2\eta_1c_R[S(-,p_2,p_3)\mu_4c'_L-S(-,p_2,p_4)\mu_3c'_R]$ \\ \hline
$+-++$ & $-2\eta_4c'_R[S(+,p_3,p_1)\mu_2c_R-S(+,p_3,p_2)\mu_1c_L]$ \\ \hline
$++--$ & $-2[S(+,p_1,p_4)S(-,p_2,p_3)c'_Lc_R
            -\mu_1\mu_2\eta_3\eta_4c'_Lc_L
            -\eta_1\eta_2\mu_3\mu_4c'_Rc_R]$  \\ \hline
$+-+-$ & $0$ \\ \hline
$+--+$ & $-2[\mu_1\mu_4\eta_2\eta_3c'_Lc_L
+\mu_2\mu_3\eta_1\eta_4c'_Rc_R
-\mu_2\mu_4\eta_1\eta_3c'_Lc_R
-\mu_1\mu_3\eta_2\eta_4c'_Rc_L]$ \\ \hline
$+---$ & $-2\eta_3c'_L[S(+,p_2,p_4)\mu_1c_L-S(+,p_1,p_4)\mu_2c_R]$ \\ \hline
\end{tabular}
\caption{The $Z$ functions for all independent helicity combinations
in terms of the functions $S$, $\eta$ and $\mu$ defined in the text.
The remaining $Z$ functions can be obtained by flipping the sign 
of the helicities
and exchanging $+$ with $-$ in the $S$ functions and $R$ with $L$ in the
chiral coefficients.}
\label{tab_Z}
\end{center}
\end{table}

In this method, all spinors for any physical momentum are defined in
terms of a basic spinor of an  auxiliary light-like 
momentum. By decomposing the
internal momenta in terms of the external ones, using Dirac algebra
and rewriting the polarisation of external vectors by means of a spinor
current, all amplitudes can eventually be 
reduced to an algebraic combination of spinors 
products $\bar u(p_i)u(p_j)$ (with $i,j,...$ labelling the external 
particles). 

\vskip.15in

{  (i)} {\it Spinors.} Since all vector particles entering our calculation
eventually splits into fermions, we only need to introduce the treatment
of the spinor fields.
External fermions\footnote{We shall use the term
`fermion' and the symbol `$u$' for both particles and antiparticles.}
 of mass $m$ and momentum $p^\mu$
are described by spinors
corresponding to states of definite helicity $\lambda$,
$u(p,\lambda)$\footnote{Here, $p$($\lambda$) represents a
generic (anti)spinor four-momentum(helicity).}, verifying the Dirac equations
\be
p\Dir u(p,\lambda)=\pm m u(p,\lambda),\qquad
\bar u(p,\lambda)p\Dir =\pm m \bar u(p,\lambda),
\ee
and the spin sum relation
\be\label{dirac}
\sum_{\lambda=\pm} u(p,\lambda)\bar u(p,\lambda)=
p\Dir\pm m,
\ee
where the sign $+(-)$ refers (here and in the following)
to a particle(antiparticle).
One can choose two arbitrary
vectors $k_0$  and $k_1$ such that
\be 
k_0\cdot k_0=0, \quad\quad k_1\cdot k_1=-1, \quad\quad k_0\cdot k_1=0,
\ee
and express the
spinors $u(p,\lambda)$ in terms of chiral ones
$w(k_0,\lambda)$ as
\be
u(p,\lambda)=w(p,\lambda)+\mu w(k_0,-\lambda),
\ee
where
\be
w(p,\lambda)=p\Dir w(k_0,-\lambda)/\eta,
\ee
and
\be
 \mu=\pm {m\over{\eta}}, \quad\quad\quad \eta=\sqrt{2|p\cdot k_0|}.
\ee
The spinors $w(k_0,\lambda)$ satisfy
\be
w(k_0,\lambda)\bar w(k_0,\lambda)=
{{1+\lambda\gamma_5}\over{2}}{k\Dir}_0,
\ee
and therefore
\be
\sum_{\lambda=\pm}w(k_0,\lambda)\bar w(k_0,\lambda)=
{k\Dir}_0.
\ee
The phase between chiral states is fixed by

\be
w(k_0,\lambda)=\lambda {k\Dir}_1 w(k_0,-\lambda).
\ee
The freedom in choosing $k_0$ and $k_1$ provides a powerful tool
for checking the correctness of any calculation.
A convenient, though not unique choice, is the following:
$k_0=(1,0,0,-1)$ and $k_1=(0,1,0,0)$.
In such a case the {\sl massless} spinors in the two methods
\cite{HZ} and \cite{KS} coincide exactly, so that it is possible 
to compare in greater detail the two corresponding numerical codes.
In particular, the results obtained with the two formalisms must agree
for every single diagram and every polarisation of external particles. 

\vskip.15in

{(ii)} {\it The $S$, $X$ and $Z$ functions.}
Using the above definitions one can compute the spinor functions
\begin{equation}
S(\lambda,p_1,p_2)=[\bar u(p_1,\lambda) u(p_2,-\lambda)],
\end{equation}
\begin{equation}
X(p_1,\lambda_1;p_2;p_3,\lambda_3;c_R,c_L)=
[\bar u(p_1,\lambda_1) {p_{2}}\Dirin\Gamma u(p_3,\lambda_3)],
\end{equation}
and
$$
\hskip -1.2in
Z(p_1,\lambda_1;p_2,\lambda_2;p_3,\lambda_3;p_4,\lambda_4;
c_R,c_L;c'_R,c'_L)= 
$$
\begin{equation}
\hskip1.7in
[\bar u(p_1,\lambda_1) \Gamma^{\mu} u(p_2,\lambda_2)]
[\bar u(p_3,\lambda_3) \Gamma'_{\mu} u(p_4,\lambda_4)],
\end{equation}
where
\begin{equation}
\Gamma^{(')\mu}=\gamma^{\mu}\Gamma^{(')},
\end{equation}
and
\begin{equation}\label{vertex}
\Gamma^{(')}=c^{(')}_R P_R + c^{(')}_L P_L,
\end{equation}
with
\begin
{equation}P_R={{1+\gamma_5}\over{2}},\quad\quad\quad
P_L={{1-\gamma_5}\over{2}},
\end{equation}
the chiral projectors. \par
By computing the resulting traces one easily finds ($\varepsilon^{0123} = 1$
is the Levi-Civita tensor) \cite{KS,mana}
\begin{equation}
S(+,p_1,p_2)= 2{{(p_1\cdot k_0)(p_2\cdot k_1)
 -(p_1\cdot k_1)(p_2\cdot k_0)
 +i\varepsilon_{\mu\nu\rho\sigma}
  k^\mu_0k^\nu_1p^\rho_1p^\sigma_2}\over{\eta_1\eta_2}},
\end{equation}
for the $S$ functions
and the expressions listed in Tabs.~\ref{tab_X} and \ref{tab_Z}
for the $X$ and $Z$ functions, respectively. For the $S$ functions, one has
$S(-,p_1,p_2)= S(+,p_2,p_1)^*$,
while the remaining $X$ and $Z$ functions can be obtained as described
in the captions of Tabs.~\ref{tab_X}--\ref{tab_Z}.

Other than the spinor parts, to each of the basic amplitudes are 
associated propagators 
functions.
In the case of off-shell fermions, they have the form
\be\label{fdenom}
D_f(\sum_{i} p_i)=\frac{1}{(\sum_{i} p_i)^2-m_f^2},
\ee
whereas in the case of bosons one gets
\be\label{bdenom}
D_V(\sum_{i} p_i)=\frac{1}{(\sum_i p_i)^2-M_V^2+iM_V\Gamma_V}.
\ee 
In eqs.~(\ref{fdenom})--(\ref{bdenom}),
$f$ is the flavour $q$, $q'$ of a virtual fermion line, whereas 
$V=W^\pm,\gamma,Z$ or $g$, being   $M_{V}=\Gamma_{V}\equiv0$ if 
$V=\gamma$ or $g$.

As for the couplings, the notation $(c_{R}^{(f)V},c_{L}^{(f)V})$ will
refer to the 
pair of chiral indices $(c_R,c_L)$ of eq.~(\ref{vertex}) entering in the 
expressions given in
Tabs.~\ref{tab_X}--\ref{tab_Z} and associated with the vertex involving 
a fermion $f=e,q,q',q''$ (whose label will only appear if the vertex is
flavour dependent) and a gauge vector $V=W^\pm,\gamma,Z,g$, according
to Tab.~\ref{tab_cRcL}. 
(Note  that we will only need using the flavour-dependent terms
$(c_{R}^{eV},c_{L}^{eV})$, i.e., for $f= e$ and $V=\gamma,Z$.)
For convenience, we also introduce the 
relative couplings $g_\gamma=1$ and $g_Z=g_\gamma/\tan\theta_W$ entering
the $\gamma W^+W^-$ and $ZW^+W^-$ vertices, respectively,
where $\theta_W$ is the Weinberg 
angle. 

\begin{table}[!h]
\begin{center}
\begin{tabular}{|c|c|c|c|c|}
\hline
\rule[0cm]{0cm}{0cm}
$~$ & $\gamma$ & $Z$ & $W^\pm$ & $g$  \\ \hline\hline
\rule[0cm]{0cm}{0cm}
$c_R$ & $Q^{f}$& $g_R^f/s_W c_W$        &  
$0$ & $1$ \\ \hline
$c_L$ & $Q^{f}$& $g_L^f/s_W c_W$ &  
$1/\sqrt{2}s_W$  &  $1$ \\ \hline
\end{tabular}
\caption{The couplings $c_R$ and $c_L$ of eq.~(\ref{vertex}) for $u$ and
$d$ type (anti)quarks and electrons/positrons
to the gauge bosons $\gamma$, $Z$, $W^\pm$ and $g$. One has
(adopting the notations $s_W\equiv \sin\theta_W$ and 
$c_W\equiv \cos\theta_W$)
$g_R^f=-Q^fs^2_W$ and $g_L^f=T_3^f-Q^fs^2_W$ (with $q=u,d$), where
$(Q^u,T^u_3)=(+\frac{2}{3},+\frac{1}{2})$, 
$(Q^d,T^d_3)=(-\frac{1}{3},-\frac{1}{2})$ and
$(Q^e,T^e_3)=(-{1},-\frac{1}{2})$
are the fermion charges and isospins.}
\label{tab_cRcL}
\end{center}
\end{table}

We are now ready to present the explicit expressions of the helicity
amplitudes associated to our process. To do so, we conventionally assume that
initial state momenta are incoming, whereas the final state ones are outgoing.
This way, we can define $-b_i=b_j=1$, where $i=1,2$ and $j=3, ...8$, so that
$\sum_{k=1,...8} b_k p_k=0$.
In correspondence to the graphs in Figs.~\ref{fig_graphs}, one can write
the Feynman amplitude squared, summed/averaged over final/initial colours
and spin, as
\begin{equation}\label{M2}
|{M}|^2=\frac{g_s^2 e^4}{4} \sum_{ \{\lambda\} }
\sum_{i=1}^{8}\sum_{j=1}^{8}
C_{ij} T_i  (\{\lambda\})
       T_j^*(\{\lambda\}), 
\end{equation}
where $C_{ij}$ are the colour factors (see below)
and with $\sum_{\{\lambda\}}$ referring to a summation over
all possible combinations of the helicities $\lambda_1, ... \lambda_8$ 
of the external particles. The quantities $g_s$ and $e$ are related to 
the aforementioned couplings by the usual relations 
$g_s^2\equiv {4\pi \alpha_s}$ and
$e^2  \equiv {4\pi \alpha_{em}}$ (in natural units). 
Assuming, for sake of illustration, that the 
process (via $W^+W^-$)
\vskip0.5cm\noindent\hskip0.75cm
\framebox{$e^+(p_1,\lambda_1)~~e^-   (p_2,\lambda_2) \longrightarrow
         ~~u  (p_3,\lambda_3)~~\bar d(p_4,\lambda_4)
         ~~s  (p_5,\lambda_5)~~\bar c(p_6,\lambda_6)
         ~~b  (p_7,\lambda_7)~~\bar b(p_8,\lambda_8)$}
\vskip0.5cm\noindent
has to be calculated, then the helicity amplitudes $T_i$ can be written as 

\begin{eqnarray}\label{T1}
T_1(\{\lambda\}) & = & - D_g(p_7+p_8)     
                         D_{W}(p_3+p_4+p_7+p_8) D_{W}(p_5+p_6)
\\ \nonumber
&        &\phantom{-}    D_u(p_3+p_7+p_8)         D_{\nu_e}(p_2-p_5-p_6)
\\ \nonumber
&        &\sum_{i=3,7,8}\sum_{j=2,5,6}b_ib_j
         {\sum_{\lambda=\pm}}{\sum_{\lambda'=\pm}}
\\ \nonumber
&        & Z(p_7,\lambda_7;p_8,-\lambda_8;p_3,\lambda_3;p_i,\lambda;
           c_{R}^g,c_{L}^g;c_{R}^g,c_{L}^g) 
\\ \nonumber
&        & Z(p_i,\lambda;p_4,-\lambda_4;p_1,\lambda_1;p_j,\lambda';
           c_{R}^W,c_{L}^W;c_{R}^W,c_{L}^W) 
\\ \nonumber
&        & Z(p_j,\lambda';p_2,-\lambda_2;p_5,\lambda_5;p_6,-\lambda_6;
           c_{R}^W,c_{L}^W;c_{R}^W,c_{L}^W) ,
\end{eqnarray}

\begin{eqnarray}\label{T2}
T_2(\{\lambda\}) & = & + D_g(p_7+p_8)     
                         D_{W}(p_3+p_4+p_7+p_8) D_{W}(p_5+p_6)
\\ \nonumber
&        &\phantom{+}    D_d(p_4+p_7+p_8)       D_{\nu_e}(p_2-p_5-p_6)
\\ \nonumber
&        &\sum_{i=4,7,8}\sum_{j=2,5,6}b_ib_j
         {\sum_{\lambda=\pm}}{\sum_{\lambda'=\pm}}
\\ \nonumber
&        & Z(p_7,\lambda_7;p_8,-\lambda_8;p_i,\lambda;p_4,-\lambda_4;
           c_{R}^g,c_{L}^g;c_{R}^g,c_{L}^g) 
\\ \nonumber
&        & Z(p_3,\lambda_3;p_i,\lambda;p_1,\lambda_1;p_j,\lambda';
           c_{R}^W,c_{L}^W;c_{R}^W,c_{L}^W) 
\\ \nonumber
&        & Z(p_j,\lambda';p_2,-\lambda_2;p_5,\lambda_5;p_6,-\lambda_6;
           c_{R}^W,c_{L}^W;c_{R}^W,c_{L}^W) ,
\end{eqnarray}

\begin{eqnarray}\label{T3}
T_3(\{\lambda\}) & = & + D_g(p_7+p_8)     
                         D_{W}(p_3+p_4)   D_{W}(p_5+p_6+p_7+p_8) 
\\ \nonumber
&        &\phantom{+}    D_s(p_5+p_7+p_8) D_{\nu_e}(p_1-p_3-p_4)
\\ \nonumber
&        &\sum_{i=5,7,8}\sum_{j=1,3,4}b_ib_j
         {\sum_{\lambda=\pm}}{\sum_{\lambda'=\pm}}
\\ \nonumber
&        & Z(p_7,\lambda_7;p_8,-\lambda_8;p_5,\lambda_5;p_i,\lambda;
           c_{R}^g,c_{L}^g;c_{R}^g,c_{L}^g) 
\\ \nonumber
&        & Z(p_i,\lambda;p_6,-\lambda_6;p_j,\lambda';p_2,-\lambda_2;
           c_{R}^W,c_{L}^W;c_{R}^W,c_{L}^W) 
\\ \nonumber
&        & Z(p_1,\lambda_1;p_j,\lambda';p_3,\lambda_3;p_4,-\lambda_4;
           c_{R}^W,c_{L}^W;c_{R}^W,c_{L}^W) ,
\end{eqnarray}

\begin{eqnarray}\label{T4}
T_4(\{\lambda\}) & = & - D_g(p_7+p_8)     
                         D_{W}(p_3+p_4)   D_{W}(p_5+p_6+p_7+p_8)
\\ \nonumber
&        &\phantom{-}    D_c(p_6+p_7+p_8) D_{\nu_e}(p_1-p_3-p_4)
\\ \nonumber
&        &\sum_{i=6,7,8}\sum_{j=1,3,4}b_ib_j
          {\sum_{\lambda=\pm}}{\sum_{\lambda'=\pm}}
\\ \nonumber
&        & Z(p_7,\lambda_7;p_8,-\lambda_8;p_i,\lambda;p_6,-\lambda_6;
           c_{R}^g,c_{L}^g;c_{R}^g,c_{L}^g) 
\\ \nonumber
&        & Z(p_5,\lambda_5;p_i,\lambda;p_j,\lambda';p_2,-\lambda_2;
           c_{R}^W,c_{L}^W;c_{R}^W,c_{L}^W) 
\\ \nonumber
&        & Z(p_1,\lambda_1;p_j,\lambda';p_3,\lambda_3;p_4,-\lambda_4;
           c_{R}^W,c_{L}^W;c_{R}^W,c_{L}^W) ,
\end{eqnarray}

\begin{eqnarray}\label{T5}
T_5(\{\lambda\}) & = & + D_g(p_7+p_8)     
                         D_{W}(p_3+p_4+p_7+p_8) D_{W}(p_5+p_6)
\\ \nonumber
&        &\phantom{+}    D_u(p_3+p_7+p_8) \sum_{V=\gamma,Z} g_V D_{V}(p_1+p_2)
\\ \nonumber
&        &\sum_{i=3,7,8}b_i
         {\sum_{\lambda=\pm}}
\\ \nonumber
&        &\{Z(p_1,\lambda_1;p_2,-\lambda_2;p_i,\lambda;p_4,-\lambda_4;
            c_{R}^{eV},c_{L}^{eV};c_{R}^W,c_{L}^W) 
\\ \nonumber
&        & \phantom{\{}
           [\sum_{j=3,4,7,8}b_j
            X(p_5,\lambda_5;p_j;p_6,-\lambda_6;c_{R}^W,c_{L}^W)-
\\ \nonumber
&        & \phantom{\{}
           \hskip0.3cm\sum_{j=1,2}b_j
            X(p_5,\lambda_5;p_j;p_6,-\lambda_6;c_{R}^W,c_{L}^W)] +
\\ \nonumber
&        & \phantom{\{}
            Z(p_i,\lambda;p_4,-\lambda_4;p_5,\lambda_5;p_6,-\lambda_6;
            c_{R}^W,c_{L}^W;c_{R}^W,c_{L}^W) 
\\ \nonumber
&        & \phantom{\{}
           [\hskip0.20cm\sum_{j=5,6}b_j
            X(p_1,\lambda_1;p_j;p_2,-\lambda_2;c_{R}^{eV},c_{L}^{eV})-
\\ \nonumber
&        & \phantom{\{}
           \sum_{j=3,4,7,8}b_j
            X(p_1,\lambda_1;p_j;p_2,-\lambda_2;c_{R}^{eV},c_{L}^{eV})] +
\\ \nonumber
&        & \phantom{\{}
            Z(p_5,\lambda_5;p_6,-\lambda_6;p_1,\lambda_1;p_2,-\lambda_2;
            c_{R}^W,c_{L}^W;c_{R}^{eV},c_{L}^{eV}) 
\\ \nonumber
&        & \phantom{\{}
           [\sum_{j=1,2}b_j
            X(p_i,\lambda;p_j;p_4,-\lambda_4;c_{R}^W,c_{L}^W)-
\\ \nonumber
&        & \phantom{\{}
           \hskip0.075cm\sum_{j=5,6}b_j
            X(p_i,\lambda;p_j;p_4,-\lambda_4;c_{R}^W,c_{L}^W)]\}
\\ \nonumber
&        & Z(p_7,\lambda_7;p_8,-\lambda_8;p_3,\lambda_3;p_i,\lambda;
           c_{R}^g,c_{L}^g;c_{R}^g,c_{L}^g) ,
\end{eqnarray}

\begin{eqnarray}\label{T6}
T_6(\{\lambda\}) & = & - D_g(p_7+p_8)     
                         D_{W}(p_3+p_4+p_7+p_8) D_{W}(p_5+p_6)
\\ \nonumber
&        &\phantom{+}    D_d(p_4+p_7+p_8) \sum_{V=\gamma,Z} g_V D_{V}(p_1+p_2)
\\ \nonumber
&        &\sum_{i=4,7,8}b_i
         {\sum_{\lambda=\pm}}
\\ \nonumber
&        &\{Z(p_1,\lambda_1;p_2,-\lambda_2;p_3,\lambda_3;p_i,\lambda;
            c_{R}^{eV},c_{L}^{eV};c_{R}^W,c_{L}^W) 
\\ \nonumber
&        & \phantom{\{}
           [\sum_{j=3,4,7,8}b_j
            X(p_5,\lambda_5;p_j;p_6,-\lambda_6;c_{R}^W,c_{L}^W)-
\\ \nonumber
&        & \phantom{\{}
           \hskip0.3cm\sum_{j=1,2}b_j
            X(p_5,\lambda_5;p_j;p_6,-\lambda_6;c_{R}^W,c_{L}^W)] +
\\ \nonumber
&        & \phantom{\{}
            Z(p_3,\lambda_3;p_i,\lambda;p_5,\lambda_5;p_6,-\lambda_6;
            c_{R}^W,c_{L}^W;c_{R}^W,c_{L}^W) 
\\ \nonumber
&        & \phantom{\{}
           [\hskip0.20cm\sum_{j=5,6}b_j
            X(p_1,\lambda_1;p_j;p_2,-\lambda_2;c_{R}^{eV},c_{L}^{eV})-
\\ \nonumber
&        & \phantom{\{}
           \sum_{j=3,4,7,8}b_j
            X(p_1,\lambda_1;p_j;p_2,-\lambda_2;c_{R}^{eV},c_{L}^{eV})] +
\\ \nonumber
&        & \phantom{\{}
            Z(p_5,\lambda_5;p_6,-\lambda_6;p_1,\lambda_1;p_2,-\lambda_2;
            c_{R}^W,c_{L}^W;c_{R}^{eV},c_{L}^{eV}) 
\\ \nonumber
&        & \phantom{\{}
           [\sum_{j=1,2}b_j
            X(p_3,\lambda_3;p_j;p_i,\lambda;c_{R}^W,c_{L}^W)-
\\ \nonumber
&        & \phantom{\{}
           \hskip0.075cm\sum_{j=5,6}b_j
            X(p_3,\lambda_3;p_j;p_i,\lambda;c_{R}^W,c_{L}^W)]\}
\\ \nonumber
&        & Z(p_7,\lambda_7;p_8,-\lambda_8;p_i,\lambda;p_4,-\lambda_4;
           c_{R}^g,c_{L}^g;c_{R}^g,c_{L}^g) ,
\end{eqnarray}

\begin{eqnarray}\label{T7}
T_7(\{\lambda\}) & = & - D_g(p_7+p_8)     
                         D_{W}(p_3+p_4) D_{W}(p_5+p_6+p_7+p_8)
\\ \nonumber
&        &\phantom{+}    D_s(p_5+p_7+p_8) \sum_{V=\gamma,Z} g_V D_{V}(p_1+p_2)
\\ \nonumber
&        &\sum_{i=5,7,8}b_i
         {\sum_{\lambda=\pm}}
\\ \nonumber
&        &\{Z(p_1,\lambda_1;p_2,-\lambda_2;p_i,\lambda;p_6,-\lambda_6;
            c_{R}^{eV},c_{L}^{eV};c_{R}^W,c_{L}^W) 
\\ \nonumber
&        & \phantom{\{}
           [\sum_{j=5,6,7,8}b_j
            X(p_3,\lambda_3;p_j;p_4,-\lambda_4;c_{R}^W,c_{L}^W)-
\\ \nonumber
&        & \phantom{\{}
           \hskip0.3cm\sum_{j=1,2}b_j
            X(p_3,\lambda_3;p_j;p_4,-\lambda_4;c_{R}^W,c_{L}^W)] +
\\ \nonumber
&        & \phantom{\{}
            Z(p_i,\lambda;p_6,-\lambda_6;p_3,\lambda_3;p_4,-\lambda_4;
            c_{R}^W,c_{L}^W;c_{R}^W,c_{L}^W) 
\\ \nonumber
&        & \phantom{\{}
           [\hskip0.20cm\sum_{j=3,4}b_j
            X(p_1,\lambda_1;p_j;p_2,-\lambda_2;c_{R}^{eV},c_{L}^{eV})-
\\ \nonumber
&        & \phantom{\{}
           \sum_{j=5,6,7,8}b_j
            X(p_1,\lambda_1;p_j;p_2,-\lambda_2;c_{R}^{eV},c_{L}^{eV})] +
\\ \nonumber
&        & \phantom{\{}
            Z(p_3,\lambda_3;p_4,-\lambda_4;p_1,\lambda_1;p_2,-\lambda_2;
            c_{R}^W,c_{L}^W;c_{R}^{eV},c_{L}^{eV}) 
\\ \nonumber
&        & \phantom{\{}
           [\sum_{j=1,2}b_j
            X(p_i,\lambda;p_j;p_6,-\lambda_6;c_{R}^W,c_{L}^W)-
\\ \nonumber
&        & \phantom{\{}
           \hskip0.075cm\sum_{j=3,4}b_j
            X(p_i,\lambda;p_j;p_6,-\lambda_6;c_{R}^W,c_{L}^W)]\}
\\ \nonumber
&        & Z(p_7,\lambda_7;p_8,-\lambda_8;p_5,\lambda_5;p_i,\lambda;
           c_{R}^g,c_{L}^g;c_{R}^g,c_{L}^g) ,
\end{eqnarray}

\begin{eqnarray}\label{T8}
T_8(\{\lambda\}) & = & + D_g(p_7+p_8)     
                         D_{W}(p_3+p_4) D_{W}(p_5+p_6+p_7+p_8)
\\ \nonumber
&        &\phantom{+}    D_c(p_6+p_7+p_8) \sum_{V=\gamma,Z} g_V D_{V}(p_1+p_2)
\\ \nonumber
&        &\sum_{i=6,7,8}b_i
         {\sum_{\lambda=\pm}}
\\ \nonumber
&        &\{Z(p_1,\lambda_1;p_2,-\lambda_2;p_5,\lambda_5;p_i,\lambda;
            c_{R}^{eV},c_{L}^{eV};c_{R}^W,c_{L}^W) 
\\ \nonumber
&        & \phantom{\{}
           [\sum_{j=5,6,7,8}b_j
            X(p_3,\lambda_3;p_j;p_4,-\lambda_4;c_{R}^W,c_{L}^W)-
\\ \nonumber
&        & \phantom{\{}
           \hskip0.3cm\sum_{j=1,2}b_j
            X(p_3,\lambda_3;p_j;p_4,-\lambda_4;c_{R}^W,c_{L}^W)] +
\\ \nonumber
&        & \phantom{\{}
            Z(p_5,\lambda_5;p_i,\lambda;p_3,\lambda_3;p_4,-\lambda_4;
            c_{R}^W,c_{L}^W;c_{R}^W,c_{L}^W) 
\\ \nonumber
&        & \phantom{\{}
           [\hskip0.20cm\sum_{j=3,4}b_j
            X(p_1,\lambda_1;p_j;p_2,-\lambda_2;c_{R}^{eV},c_{L}^{eV})-
\\ \nonumber
&        & \phantom{\{}
           \sum_{j=5,6,7,8}b_j
            X(p_1,\lambda_1;p_j;p_2,-\lambda_2;c_{R}^{eV},c_{L}^{eV})] +
\\ \nonumber
&        & \phantom{\{}
            Z(p_3,\lambda_3;p_4,-\lambda_4;p_1,\lambda_1;p_2,-\lambda_2;
            c_{R}^W,c_{L}^W;c_{R}^{eV},c_{L}^{eV}) 
\\ \nonumber
&        & \phantom{\{}
           [\sum_{j=1,2}b_j
            X(p_5,\lambda_5;p_j;p_i,\lambda;c_{R}^W,c_{L}^W)-
\\ \nonumber
&        & \phantom{\{}
           \hskip0.075cm\sum_{j=3,4}b_j
            X(p_5,\lambda_5;p_j;p_i,\lambda;c_{R}^W,c_{L}^W)]\}
\\ \nonumber
&        & Z(p_7,\lambda_7;p_8,-\lambda_8;p_i,\lambda;p_6,-\lambda_6;
           c_{R}^g,c_{L}^g;c_{R}^g,c_{L}^g) .
\end{eqnarray}

Concerning the colour factors, there are basically only two of these, if the 
flavours in the final state are all different. They are (hereafter, $N_C=3$):
$C_{ij}=\frac{N_C}{4}(N_C^2-1)=6$, if $i$ and $j$ are diagrams in which
the gluon emission takes place from the same $W^\pm$, and $C_{ij}=0$
otherwise (because of colour conservation)\footnote{In other terms, 
`perturbative' colour-rearrangement is not possible in $W^+W^-\ar 6$ quarks
at ${\cal O}(\alpha_s^2)$, 
contrary to the case of four-quark-two-gluon decays at the same
order \cite{Torino}.}.

If one or two quark flavours in the final state are identical, then
the number of Feynman graphs proliferates, as explained in 
Sect.~\ref{sec_motiv}. 
The spinor part of the additional diagrams can easily be obtained by 
interchanging the labels of identical particles in the above
formulae, factorising a minus sign for each of these operations.
In addition, as many factors of the form $\frac{1}{2^n}$ multiply the 
amplitude squared as the number of $n$-tuple of identical final state
particles.
As for the new colour factors, one has to do some more work. However, it is
rather trivial to realise than only the following colour structures need
to be computed\footnote{The product of the first two yields
the factor $C_{ij}=6$ mentioned above
whereas the third produces the other one, $C_{ij}=0$.}:
\begin{picture}(600,150)
\SetScale{1.0}
\SetWidth{1.2}
\CArc(100,100)(30,270,90)
\CArc(100,100)(30,90,270)
\Text(165,100)[]{= $N_C$ = 3}
\end{picture}
\vskip-2.0cm
\begin{picture}(600,150)
\SetScale{1.0}
\SetWidth{1.2}
\CArc(100,100)(30,270,90)
\CArc(100,100)(30,90,270)
\Gluon(122,120)(178,120){3}{5}
\Gluon(178, 80)(122, 80){3}{5}
\CArc(200,100)(30,270,90)
\CArc(200,100)(30,90,270)
\Text(282.5,100)[]{= $({N_C^2-1})/{4}$ = 2}
\end{picture}
\vskip-2.0cm
\begin{picture}(600,150)
\SetScale{1.0}
\SetWidth{1.2}
\CArc(100,100)(30,270,90)
\CArc(100,100)(30,90,270)
\Gluon(130,100)(170,100){3}{4}
\CArc(200,100)(30,270,90)
\CArc(200,100)(30,90,270)
\Gluon(230,100)(270,100){3}{4}
\CArc(300,100)(30,270,90)
\CArc(300,100)(30,90,270)
\Text(347.5,100)[]{= 0}
\end{picture}
\vskip-2.0cm
\begin{picture}(600,150)
\SetScale{1.0}
\SetWidth{1.2}
\CArc(100,100)(30,270,90)
\CArc(100,100)(30,90,270)
\Gluon( 70,100)(130,100){3}{5}
\Gluon(100,130)(100, 70){3}{5}
\Text(205,100)[]{= $({1/N_C-N_C})/{4}$ = $-2/3$}
\end{picture}
\vskip-2.0cm
\begin{picture}(600,150)
\SetScale{1.0}
\SetWidth{1.2}
\CArc(100,100)(30,270,90)
\CArc(100,100)(30,90,270)
\Gluon( 85,125)( 85, 75){3}{5}
\Gluon(115, 75)(115,125){3}{5}
\Text(220,100)[]{= $({1/N_C-2N_C+N_C^3})/{4}$ = $16/3$}
\end{picture}
\vskip-2.0cm
\begin{picture}(600,150)
\SetScale{1.0}
\SetWidth{1.2}
\CArc(100,100)(30,270,90)
\CArc(100,100)(30,90,270)
\Gluon(130,100)(170,100){3}{4}
\CArc(200,100)(30,270,90)
\CArc(200,100)(30,90,270)
\Gluon(200,130)(200, 70){3}{5}
\Text(247.5,100)[]{= 0}
\end{picture}
\vskip-1.0cm
\noindent
Since the all 
procedure is quite cumbersome, we refrain here from 
building up explicitly
the correct MEs for identical flavours by combining the above 
colour factors (and their products) 
with the appropriate interferences among spinor amplitudes.
Instead, we make available  upon request our 
programs, that do include the described implementation.

Before proceeding further, we would now like to devote some space to describe
the procedure adopted to integrate the squared amplitude in eq.~(\ref{M2}).
In fact, in order to deal numerically with the non-trivial resonant structure 
of our six-quark process, one has to apply some special care. Here, 
we have adopted the technique of splitting the ME
in a sum of non-gauge-invariant pieces, each of these implementing a
different resonant structure, and of integrating them separately with
the appropriate mapping of the phase space variables.

Things go as follows. Firstly, one isolates the diagrams with similar 
resonant structure by grouping these together in `subamplitudes'. From
the graphs in Fig.~\ref{fig_graphs}, one can recognise the following 
two resonant structures: say, (a) $W^+\ar (3478)$ and $W^-\ar (56)$ 
(graphs 1,2,5,6, so that $T_a=\sum_{i=1,2,5,6}T_i$);
(b) $W^+\ar (34)$ and $W^-\ar (5678)$ (graphs 3,4,7,8,
so that $T_b=\sum_{i=3,4,7,8}T_i$).
Secondly, one defines the mentioned non-gauge-invariant components of
the amplitude squared, by appropriately combining the subamplitudes.
For example, we have simply taken the square of the two resonant subamplitudes
and their interference: $|T_a|^2$, $|T_b|^2$ and $2~\mbox{Real}(T_aT^*_b)$,
respectively. Thirdly, one maps the phase space around the resonances. 
Fourthly, the various amplitude squared terms are integrated separately
and added up in the end (to recover gauge invariance) to produce total
and differential cross sections.
In this respect, we would like to mention that all results presented here 
have been
obtained by resorting to the adaptive multi-dimensional integrator {\tt VEGAS}
\cite{VEGAS}, and they have been counter-checked against the outputs of the
multi-particle phase space generator {\tt RAMBO} \cite{RAMBO}.

\section{Results}
\label{sec_results}

In order to select a six-`jet' sample we apply a jet clustering algorithm
directly to the `quarks' in the final state of
$e^+e^-\ar W^+W^-\ar q\bar q q'\bar q' q''\bar q''$. 
For illustrative purposes, we use the Durham jet-finder 
\cite{DURHAM} only. However, 
we remark that none of the main features of our analysis 
depends drastically on such a choice. This algorithm is based 
on the `(squared) transverse-momentum' measure
\begin{equation}\label{D}
y_{ij} = {{2\min (E^2_i, E^2_j)(1-\cos\theta_{ij})}\over{s}},
\end{equation}
where $E_i$ and $E_j$ are the energies and $\theta_{ij}$ the separation
of any pair $ij$ of particles in the final state, with 
$3\le i <j=4, ... 8$, to be 
compared against a resolution parameter denoted by $y_{\mathrm{cut}}$. 
In our tree-level studies, the selected rate is then nothing else than the 
total partonic
cross section with a cut $y_{ij}>y_{\mathrm{cut}}$ on any possible $ij$ 
combination.

\begin{figure}[p]
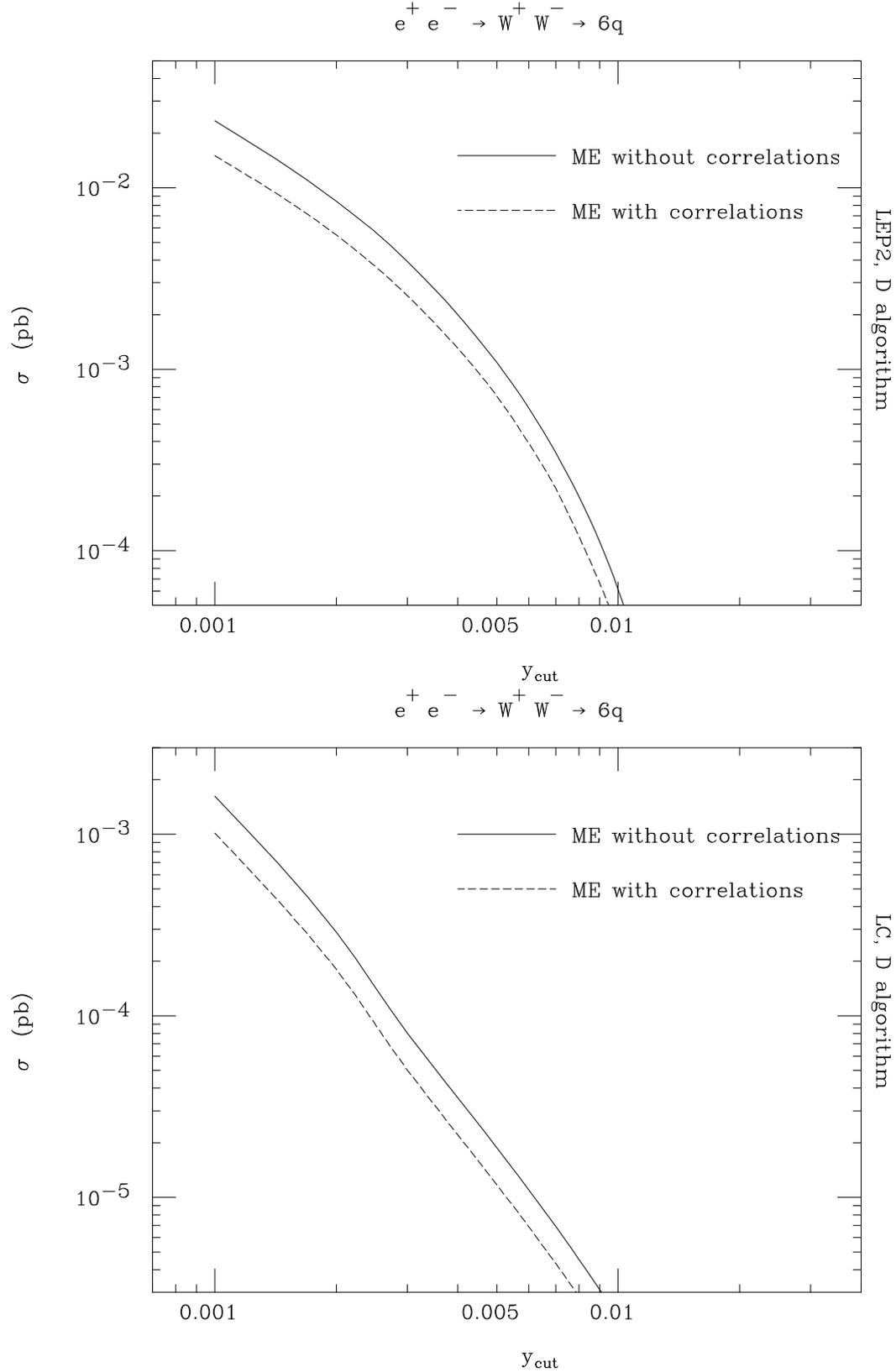

~\epsfig{file=rmt_fermidirac_rate.ps,height=14cm,angle=90}
\vskip0.0001mm
~\epsfig{file=rmt_fermidirac_rate_350.ps,height=14cm,angle=90}
\caption{Cross section for $e^+e^-\ar W^+W^-\ar q\bar q
q'\bar q' q''\bar q''$ events as a function of the resolution parameter 
$\ycut$ in the Durham jet-finder, at $\Ecm=172$ (above) and 350 (below) GeV.
The summation over all possible combinations of flavours $q$, $q'$ and $q''$ 
has been performed. Solid: without Fermi-Dirac correlations. Dashed:
with Fermi-Dirac correlations.}
\label{fig_FD_rate}
\end{figure}

Fig.~\ref{fig_FD_rate} presents the 
total cross section at LEP2 (above) and a LC (below) for six-quark events
as a function of the resolution, 
with and without the correlations described in Sect.~\ref{sec_motiv}.
(A summation over 
all possible combinations of quark flavours has been performed.)
We see a large effect on the 
integrated rates. Indeed,  the ratio between the two curves
is about 1.6 for $y_{\mathrm{cut}}\Ord0.01$ at both energies. The 
cross section including the interference effects is indeed smaller,
in accordance with the fact that these are generally destructive. 

Six-quark decays of $W^+W^-$ pairs are detectable but not numerous
at $\Ecm=172$ GeV
for $y_{\mathrm{cut}}\Ord0.004$, assuming 500 inverse picobarn of
luminosity. At the minimum $y_{\mathrm{cut}}$ considered here and
for the mentioned figure of $\int{\cal L}dt$ some 8 events should
be expected. In contrast, at a LC running at the top-antitop threshold, i.e.,  
$\Ecm\approx2m_t\approx 350$ GeV, assuming 100 to 500 fb$^{-1}$
per year (e.g., in the TESLA design), one gets between 100 and 500 events, for 
$y_{\mathrm{cut}}=0.001$. Presumably, a QCD $K$-factor of order 1.5--2
should apply  to the total production rates of
$e^+e^-\ar W^+W^-\ar q\bar q q'\bar q' q''\bar q''$, in line with the results 
obtained for $e^+e^-\ar \gamma^*,Z\ar q\bar q q'\bar q'$ \cite{a3},
so that the actual number of events detected should accordingly be larger.

\begin{figure}[p]
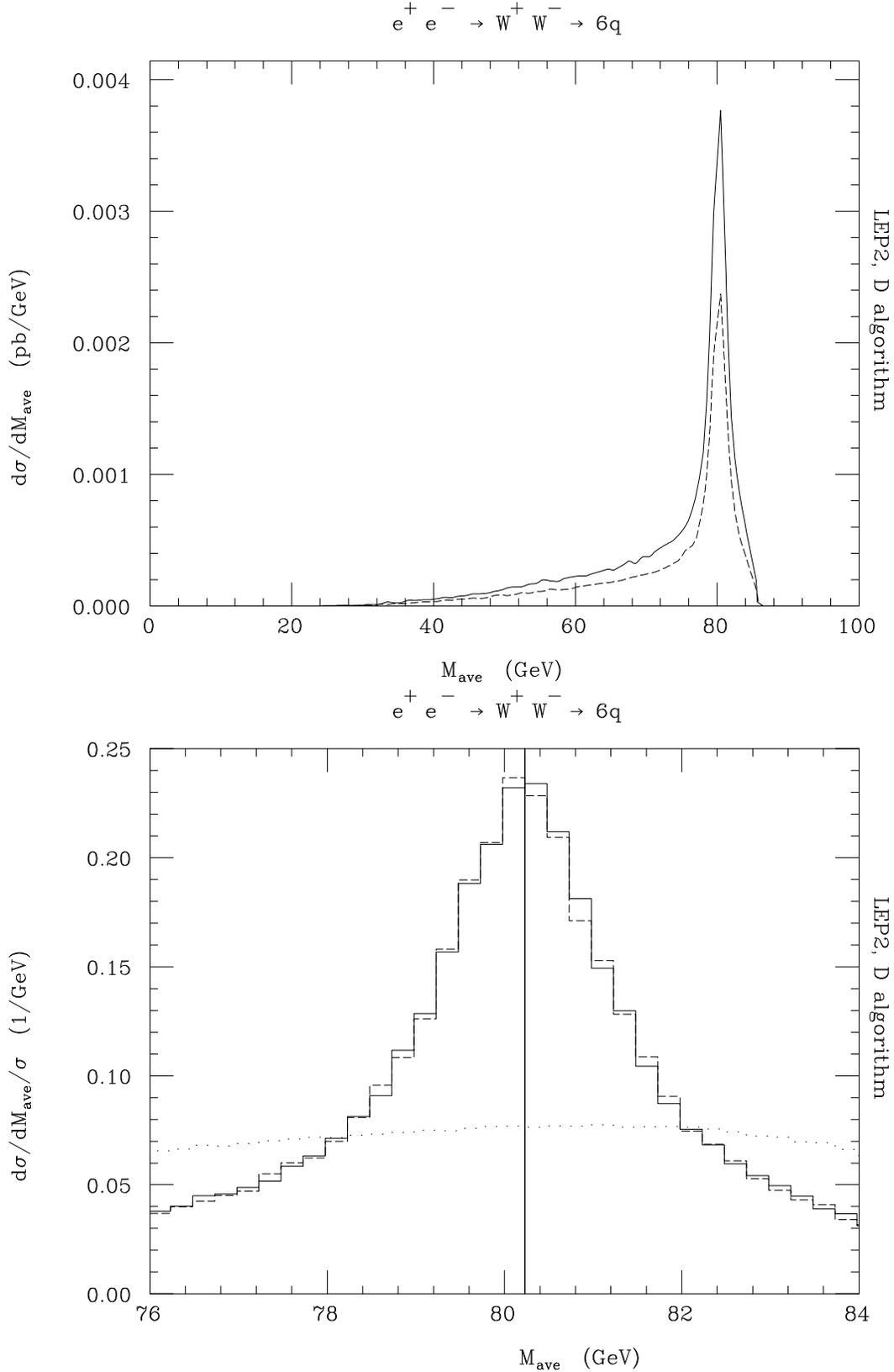

~\epsfig{file=rmt_fermidirac_dist.ps,height=14cm,angle=90}
\vskip0.0001mm
~\epsfig{file=rmt_fermidirac_blow.ps,height=14cm,angle=90}
\caption{Differential distributions in the `average value' of
the best reconstructed $W^\pm$ masses for $e^+e^-\ar W^+W^-\ar q\bar q
q'\bar q' q''\bar q''$ events (solid and dashed lines) and a for 
pure phase-space model of six-particle production (dotted line, below
only),
with $\ycut=0.001$ in the Durham jet-finder, at $\Ecm=172$ 
GeV. The summation over all possible combinations of flavours $q$, $q'$ and 
$q''$ has been performed. Solid: without Fermi-Dirac correlations. Dashed:
with Fermi-Dirac correlations.
Normalisation is to the total cross section(unity) in the plot above(below).}
\label{fig_FD_dist}
\end{figure}

As for effects of $W^+W^-\to6$~quark events onto
the line-shape of the $W^\pm$ mass resonance, we have found these
negligible. We have performed  
 {\tt MINUIT} \cite{minuit}  fits of the form
\begin{equation}\label{fitf}
f(m)=c_1\frac{c_2^2 c_3^2}{(m^2-c_2^2)^2+c_2^2 c_3^2}+g(m),
\end{equation}
where the term $g(m)$ is meant to simulate a smooth background
due to mis-assigned jets induced by the jet-clustering algorithm,
\begin{eqnarray}\label{fitg}
g(m) = \left \{ \begin{array}{c}  0, \\[3mm]
                                  c_4+c_5~(m-c_2)+c_6~(m-c_2)^2, \\[3mm]
                                  c_4\frac{1}{1+{\mathrm{exp}}((m-c_5)/c_6)},
\end{array} \right.
\end{eqnarray}
that is, a null, a three-term polynomial and a smeared
step function, see Ref.~\cite{schemes}.
Clearly, in eq.~(\ref{fitf}), we have assumed a Breit-Wigner shape
characterised by a peak height $c_1$, a position $c_2$ and a width $c_3$,
these corresponding to the normalisation, $M_W$ and $\Gamma_W$, respectively.
By adopting various algorithms and resolutions, we have never found
a difference larger than 10 MeV between the $c_2$ coefficients obtained from 
 various pairs of mass spectra $m$ (one computed with and the other
without the mentioned correlations), neither at LEP2 nor at
a LC.  Fig.~\ref{fig_FD_dist} illustrates
typical differences, e.g., in the case of an `average mass'. 
This can be obtained by applying the
Durham algorithm with $y_{\mathrm{cut}}=0.001$ to the $2\ar6$ process
$e^+e^-\ar W^+W^-\ar q\bar q q'\bar q' q''\bar q''$ and then forcing the
six-body final state into a four-body one, by clustering the 
three softest 
particles $i,j,k$ into one pseudo-particle $l$ with four-momentum
$p_l^\mu=p_i^\mu+p_j^\mu+p_k^\mu$ 
(the so-called `E recombination scheme' \cite{schemes}). 
Having done this, one looks at the 
three possible pairs of di-jet combinations
that can be formed out of the four surviving four-momenta, 
rejects the one in which the two
most energetic particles are put together 
and plots the average of the other two,
$M_{\mathrm{ave}}$.

\begin{figure}[p]
~\epsfig{file=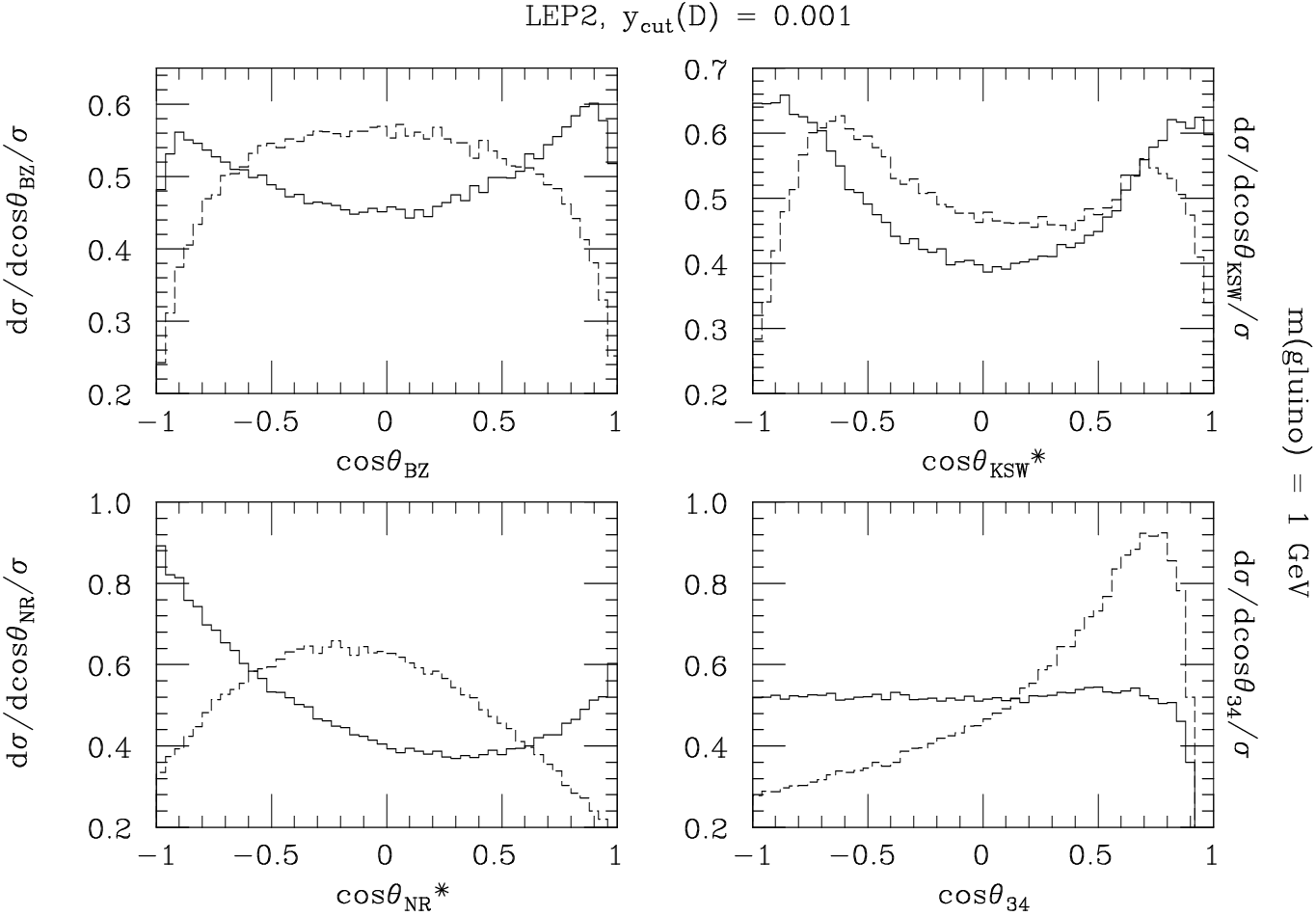,height=16cm,angle=90}
\caption{Differential distributions in the cosine of the
following angular variables:
(top-left) $\chi_{\mathrm{BZ}}$, (top-right) $\Phi_{\mathrm{KSW}}^*$,
(bottom-left) $\theta_{\mathrm{NR}}^*$ and
(bottom-right) $\theta_{34}$, for $e^+e^-\ar W^+W^-\ar q\bar q
q'\bar q' gg$ (solid lines) and $e^+e^-\ar W^+W^-\ar q\bar q
q'\bar q' \gluino\gluino$ (dashed lines) events, for $\mgluino=1$ GeV,
with $\ycut=0.001$ in the Durham jet-finder, at $\Ecm=172$ GeV. 
The summation over all possible combinations of flavours $q$ and $q'$ 
has been performed. Spectra are normalised to unity.}
\label{fig_angles_cfr}
\end{figure}

We will now proceed to studying
 the relevance of $e^+e^-\ar W^+W^-\ar q\bar 
q q'\bar q'
q''\bar q''$ events 
in the search for new particles at LEP2 and a
LC. In doing so, we will use the full ME, with all mentioned correlations
included.
For a start, we have found it rather easy 
to individuate four particle momenta out of the original six in the final
state that preserve the typical differences between 
(on the one hand) gluons and (on the other hand) quarks and gluinos in the
 variables which are used to fit the QCD colour factors
(see Footnote 1). 
This can be verified by referring to Fig.~\ref{fig_angles_cfr}, 
where the differential
distributions (normalised to unity) in the four 
variables\footnote{See, e.g.,  Ref.~\cite{ioebas} for 
the definition of the angles and for some typical spectra 
in the case of $e^+e^-\ar 4$ parton processes at the $Z$ peak.}
\begin{enumerate}
\item the Bengtsson-Zerwas angle $\chi_{\mathrm{BZ}}$;
\item the (modified) K\"orner-Schierholz-Willrodt angle 
$\Phi_{\mathrm{KSW}}^*$;
\item the (modified) Nachtmann-Reiter angle $\theta_{\mathrm{NR}}^*$;
\item the angle between the two least energetic jets $\theta_{34}$;
\end{enumerate}
have been plotted for 
$e^+e^-\ar W^+W^-\ar q\bar q q'\bar q' gg$ and 
$e^+e^-\ar W^+W^-\ar q\bar q q'\bar q' \gluino \gluino$ events, e.g.,  
at LEP2\footnote{Those
for $e^+e^-\ar W^+W^-\ar q\bar q q'\bar q' q''\bar q''$ 
coincide within numerical
errors with the latter.}. 
Here, for reference, the gluino mass has been set equal to one
(usual jet-finder and resolution have been used). 
The above four-particle
quantities have been built by simply using the four three-momenta that
survive after having removed the pair which yields the 
invariant mass closer to $M_W$. 

\begin{figure}[p]
~\epsfig{file=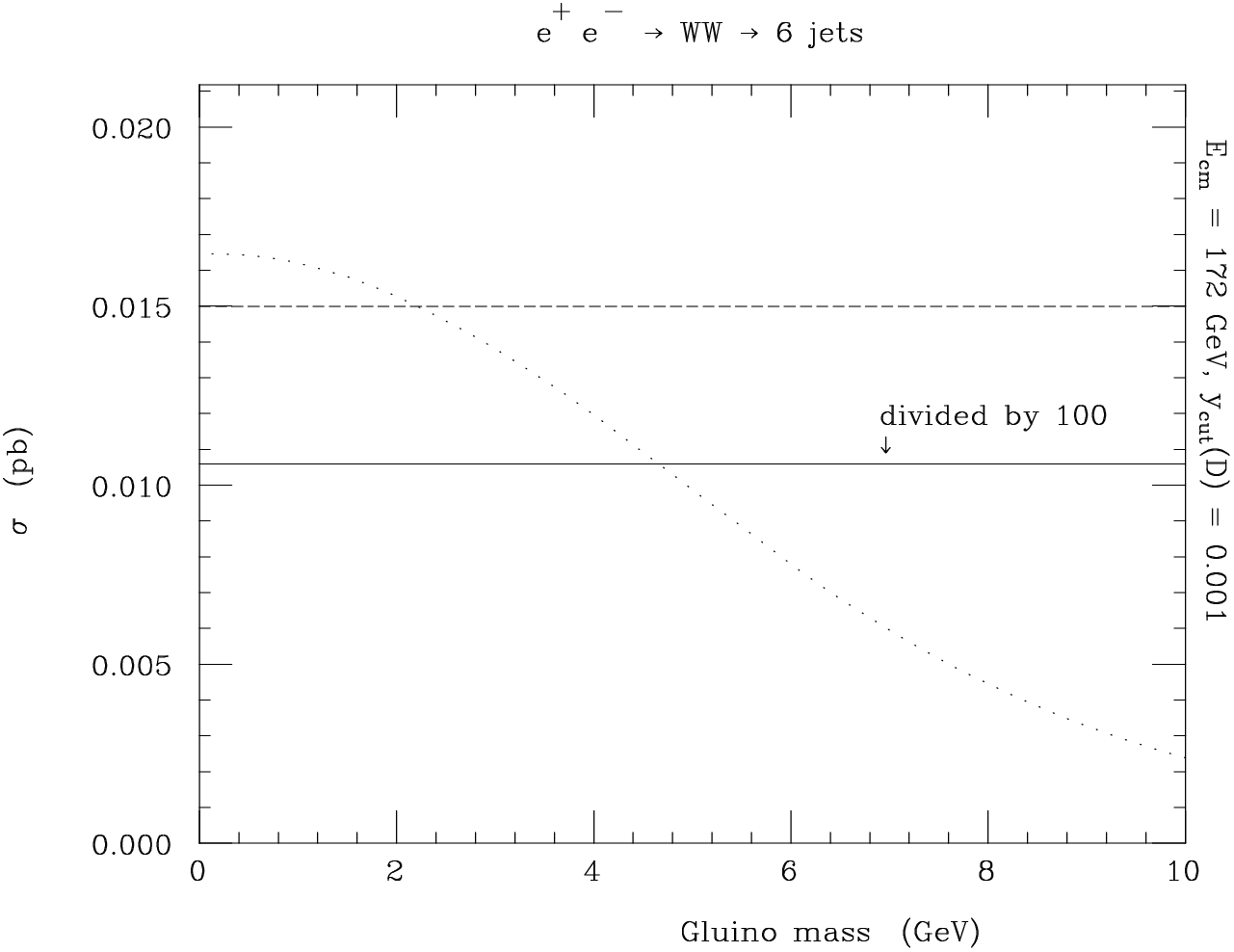,height=14cm,angle=90}
\vskip0.0001mm
~\epsfig{file=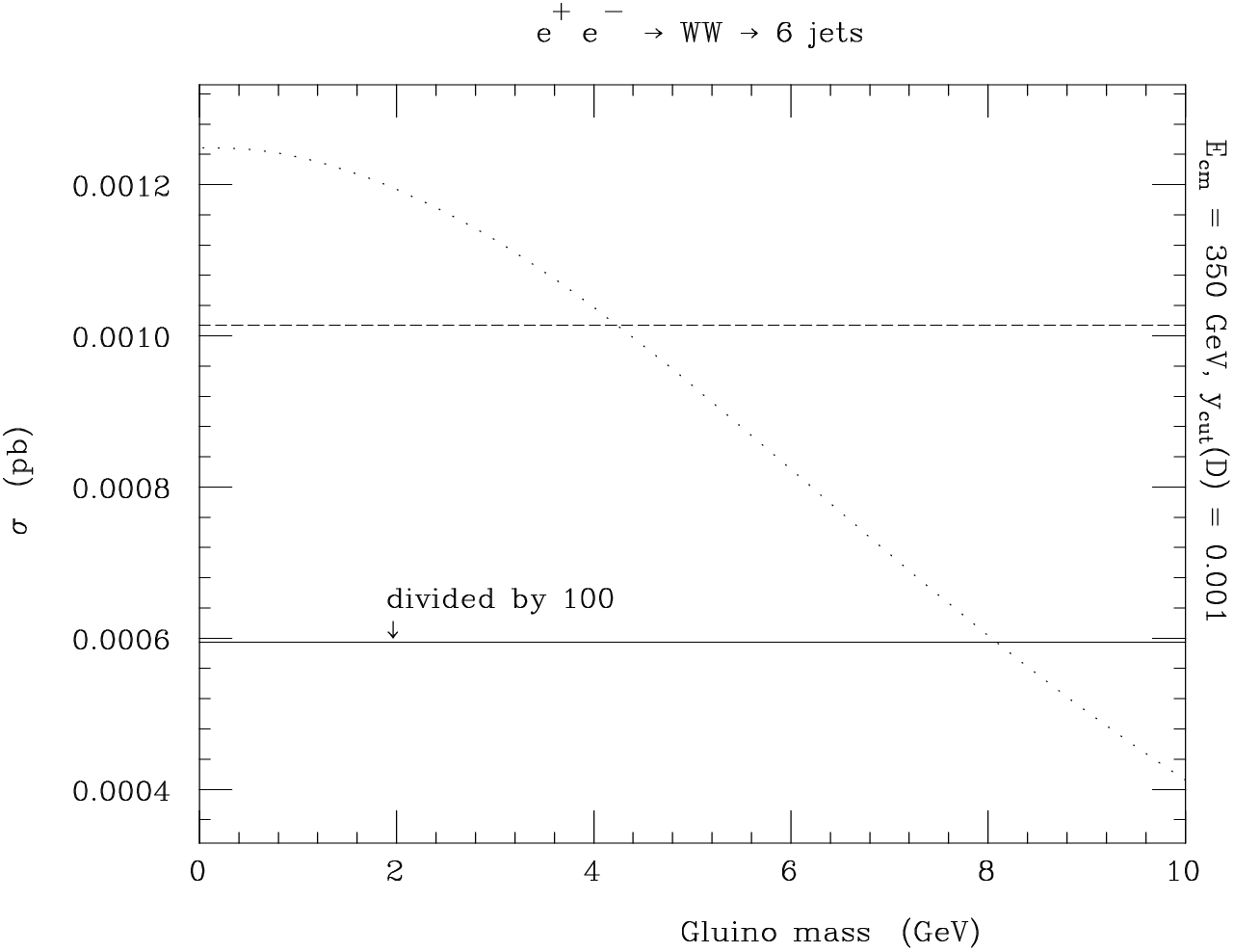,height=14cm,angle=90}
\caption{Cross section for $e^+e^-\ar W^+W^-\ar q\bar q
q'\bar q' gg$ (solid lines), $e^+e^-\ar W^+W^-\ar q\bar q
q'\bar q' q'' \bar q''$ (dashed lines) 
 and $e^+e^-\ar W^+W^-\ar q\bar q
q'\bar q' \gluino\gluino$ (dotted lines) events as a function of the
gluino mass,
with $\ycut=0.001$ in the Durham jet-finder, at $\Ecm=172$
(upper plot) and 350 GeV (lower plot). 
The summation over all possible combinations of flavours $q$ and $q'$ 
has been performed. Six-quark rates include Fermi-Dirac correlations.
Gluon rates have been divided by hundred for readability.} 
\label{fig_gluino}
\end{figure}

As for the total rates of the various six-jet contributions, these
can be found in Fig.~\ref{fig_gluino}.
 The CM energies are the same as in Fig.~\ref{fig_FD_rate}. 
It is clear from the LEP2 plot that at such a collider
there is no chance of selecting a statistically significant sample of 
SUSY events. 
 If one assumes, say, 500 inverse picobarn to be collected at LEP2,
then, from looking at the upper part of Fig.~\ref{fig_gluino}, it follows that
the number of SUSY events produced should be between 1 and 8, 
for $y_{\mathrm{cut}}=0.001$
in the Durham algorithm, depending on the actual value of the gluino mass !  
In contrast, one should expect a LC to be an
excellent laboratory for gluino searches in $W^+W^-$ decays.
(Incidentally, notice that at a LC the cross section for
$W^+W^-$ production is even larger than that for the $Z$.)
In fact, although the LC 
production rate of $e^+e^-\ar W^+W^-$ is smaller
than the LEP2 one, the instantaneous luminosity 
is in contrast much higher, since 
some 100 to 500 fb$^{-1}$ of data  per annum
are expected to be collected. 
If one goes back to the lower part of Fig.~\ref{fig_gluino} and
considers a LC running at 350 GeV, then one should expect to produce
approximately 120 to 600 events per year with very light gluinos, 
for $y_{\rm{cut}}=0.001$ in the Durham scheme (possibly, twice as much,
accounting for the $K$-factor).

\begin{figure}[p]
~\epsfig{file=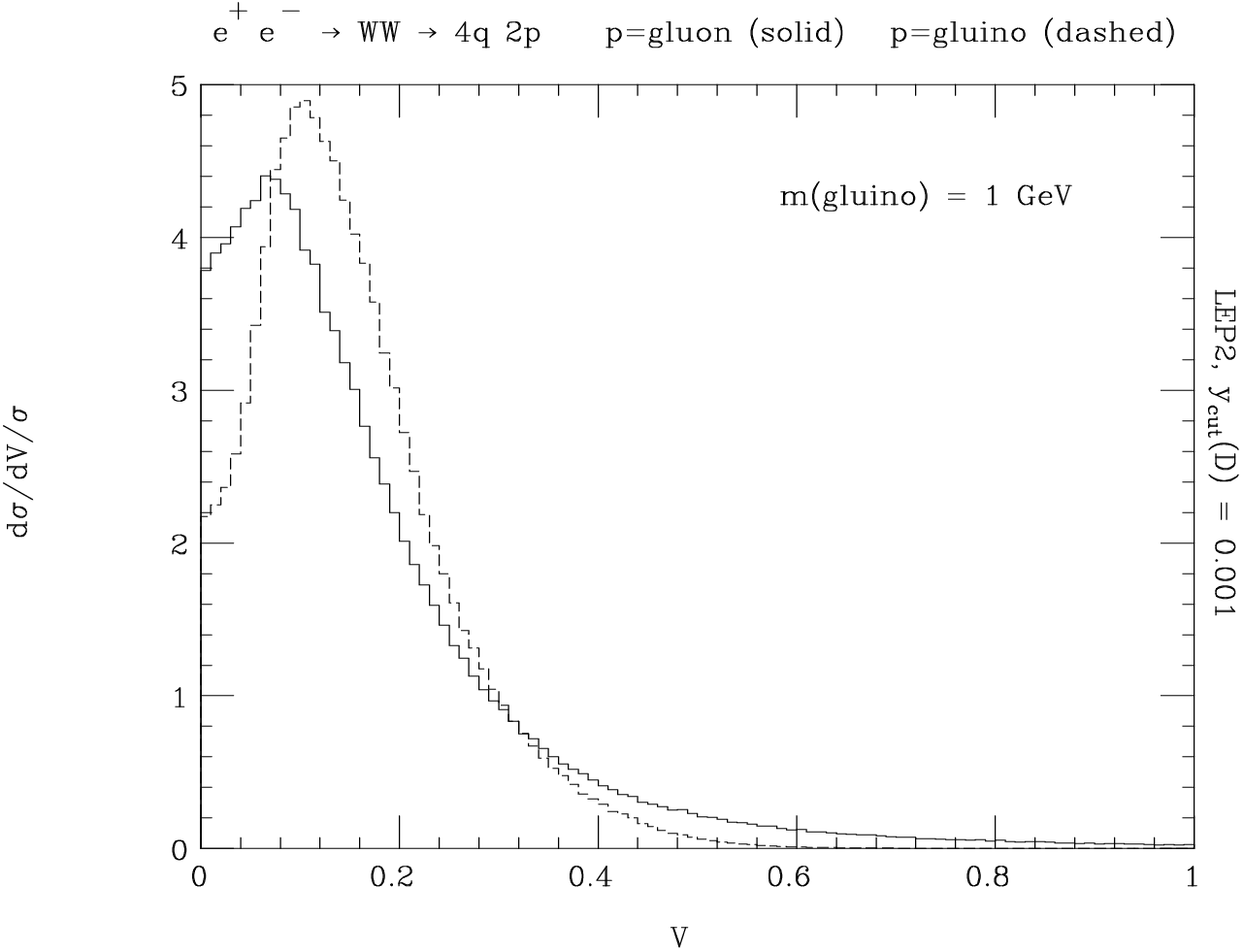,height=16cm,angle=90}
\caption{Differential distributions in the quantity defined
in eq.~(\ref{volume}) of the resonant four-jet 
subsystem for $e^+e^-\ar W^+W^-\ar q\bar q
q'\bar q' gg$ (solid lines) and $e^+e^-\ar W^+W^-\ar q\bar q
q'\bar q' \gluino\gluino$ (dashed lines) events, for $\mgluino=1$ GeV,
with $\ycut=0.001$ in the Durham jet-finder, at $\Ecm=172$ GeV. 
The summation over all possible combinations of flavours $q$ and $q'$ 
has been performed. Spectra are normalised to unity. }
\label{fig_volume}
\end{figure}

Varying the CM energy of the colliding $e^+e^-$ beams has little effects on
the effectiveness of our simple procedure based on the 
$M_W$ selection of four four-momenta. This can be seen by defining
 the Lorentz-invariant (contrary to angles) quantity
\begin{equation}\label{volume}
V=N_V\frac{\varepsilon(p_1,p_2,p_3,p_4)}{s^2},
\end{equation}
where the numerator represents the contraction of the jet four-momenta
$p_1^\mu,p_2^\nu,p_3^\rho,p_4^\sigma$ used in Fig.~\ref{fig_angles_cfr}
with the Levi-Civita tensor
$\varepsilon_{\mu\nu\rho\sigma}$ of Sect.~\ref{sec_calc} and where the
factor $N_V$ has been introduced for scaling purposes.
The quantity in eq.~(\ref{volume}) is presented
in Fig.~\ref{fig_volume}, e.g., for
 $E_{\rm{cm}}=172$ GeV, using $N_V=1000$: 
compare to Fig.~\ref{fig_angles_cfr}, where the same energy
was used. In this case, we again have looked at the 
 four-quark-two-gluon and four-quark-two-gluino
final states only, as  the six-quark contribution behaves
rather similarly to the SUSY one. Basically, $V$ is a measure
of the acoplanarity of the event \cite{acoplanarity}, this
in turn quantifying the relative orientation of the planes
spanned by, on the one hand, the two most energetic particles and, on the other
hand, the two least energetic ones. In fact, it should be recalled 
that the first
three angles introduced above are nothing else than a different way of 
describing
the helicity property that in a $g^*\ar gg$ splitting the two gluons (they are
spin 1 bosons) tend to lie in the same
plane of the two quarks which originally emitted the virtual gluon, whereas
in $g^*\ar q\bar q$ and $g^*\ar \gluino \gluino$ splittings the two 
quarks and the
two gluinos (they are both spin 1/2 fermions) tend to be in a perpendicular 
one. 
Fig.~\ref{fig_volume} eloquently confirms this dynamics.

Under these circumstances then, dedicated analyses in the angular variables
of Fig.~\ref{fig_angles_cfr} could well be attempted at a LC (also
notice the somewhat improved `SUSY to ordinary QCD' 
production ratio respect to LEP2). 
To select a six-jet sample from $W^+W^-$ decays
 should be rather straightforward, we believe, by removing
those where none of the di-jet invariant masses reproduce $M_W$,
most of which would come from ${\cal O}(\alpha_s^4)$ QCD events \cite{ee6j},
and those in which one or more three-jet masses reconstruct $m_t=175$ GeV, 
as it occurs in single- and double-top events \cite{eett}. 
In principle, one could also resort to semileptonic decays, i.e.,
$W^+W^-\to 2~{\mathrm{jets}}~\ell^\pm$ plus missing energy. 
In practise, though, the loss of kinematic constraints, because of the
neutrino escaping detection, would render the background suppression
less effective. Furthermore, to exploit the sort of
vertex tagging procedure described in Ref.~\cite{noigluino} and recalled in 
Sect.~\ref{sec_motiv} could be even more 
fruitful, given the high efficiency and purity foreseen for such a technique
by the time the electron-positron linear colliders will have started operation,
but provided that 
$\tau_\gluino\sim\frac{M^4_{\tilde q}}{\alpha_{em}\alpha_s
m_\gluino^5}\OOrd\tau_b$ (e.g., in the SUGRA scenario),
with $\tau_b\sim 1.6\cdot10^{-12}$ sec 
and where $M_{\tilde q}$ is the typical squark mass.

We now turn our attention to the case of Higgs 
searches in the $bb$ + 4 jet channel, 
at a LC\footnote{Notice that in all forthcoming
plots we have not included a multiplicative factor $\epsilon_b^2$, accounting
for the finite efficiency of tagging the two heavy quarks. We assume $\epsilon_b$
to be large enough so that the reducible background from 
$e^+e^-\ar W^+W^-\ar q\bar q q'\bar q' gg$ studied in Ref.~\cite{Torino} 
and in the first part of this Section can easily be filtered out of the 
$bb$ + 4 jet sample.}.
Though $e^+e^-\ar W^+W^-\ar  q\bar q q'\bar q' b\bar b $ events 
can be relevant as background processes in
 both the SM and the MSSM, for reasons of space, we illustrate here the phenomenology
of the latter model only. The former case can easily be dealt with by the
reader itself, by referring to the specialised bibliography on the subject 
\cite{ee500Higgs}.

Fig.~\ref{fig_higgs} presents the Higgs rates at a LC with 
$E_{\mathrm{cm}}=350$ (above) and 500 (below) GeV, for two reference values
of $\tan\beta$, 3.0 and 30.,
as a function of the scalar Higgs masses. The 
background considered
here clearly does not depend on either of them, so it is simply indicated by an arrow
in both plots, in correspondence of the transition value of the Higgs
mass between the light, $M_h$, and heavy, $M_H$, regime.
As usual, six-jet final states (here, with two $b$ quarks) are selected using
the Durham jet-finder with cut-off $y_{\mathrm{cut}}=0.001$. 
For such 
jet selection enforced, the background overwhelms both scalar Higgs signals,
$e^+e^-\ar \Phi Z$ and $e^+e^-\ar \Phi A$, over a large interval
in $M_{{h}}$ 
and $M_{{H}}$. 

\begin{figure}[p]
~\epsfig{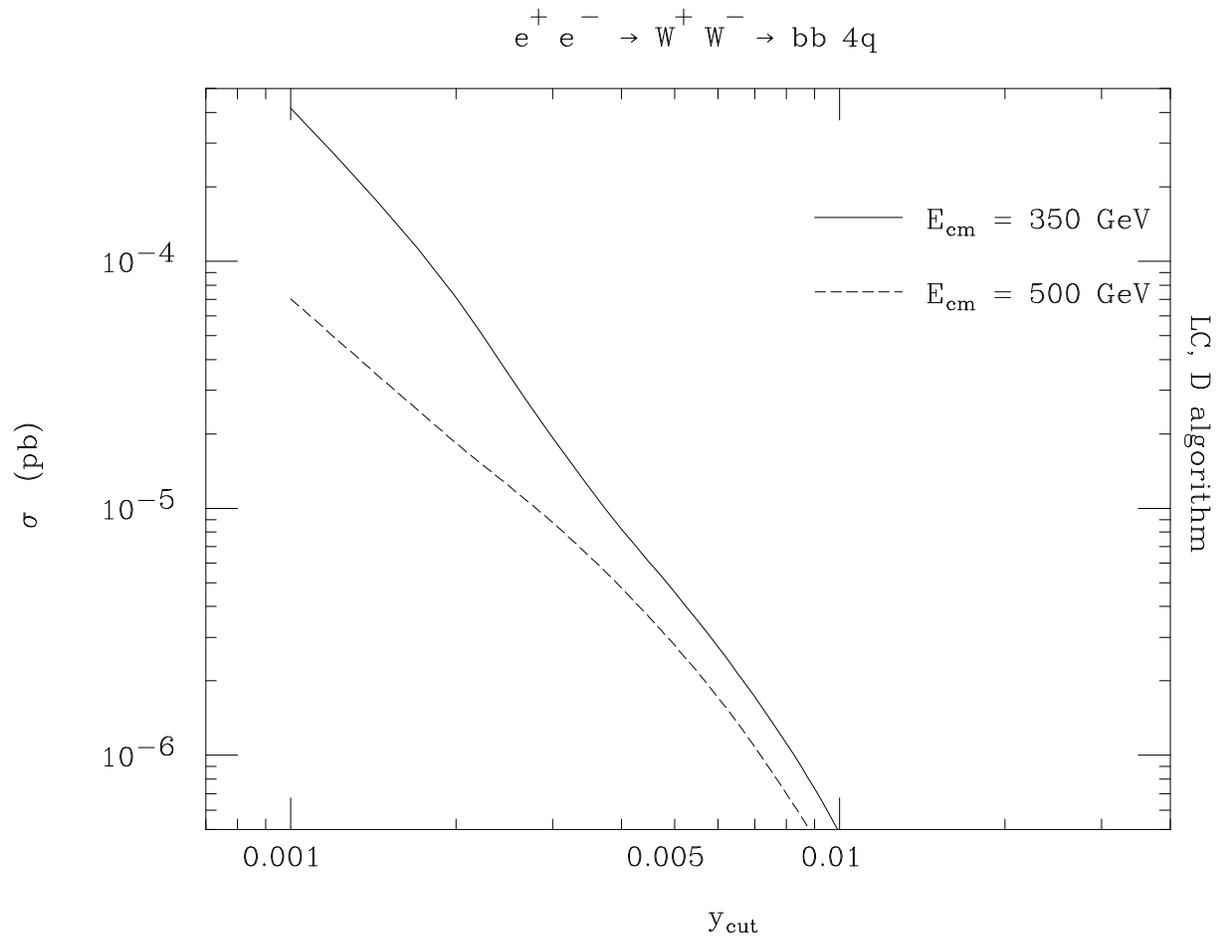}
\caption{Cross section for $e^+e^-\ar W^+W^-\ar q\bar q
q'\bar q' b\bar b $ events as a function of the resolution parameter 
$\ycut$ in the Durham jet-finder, at $\Ecm=350$ (solid line)
and 500 (dashed line) GeV.
The summation over all possible combinations of flavours $q$ and $q'$ 
has been performed. }
\label{fig_6j_nlc}
\end{figure}

However, the situation is in reality less dramatic than it would appear from 
Fig.~\ref{fig_higgs} only, if one refers to Fig.~\ref{fig_6j_nlc} too.
In fact, whereas all the decay products emerging from the $W^+W^- Z$ and 
$W^+W^- A$
intermediate states of the Higgs signals are naturally energetic and far apart,
the $b$ quarks generated in $W^+W^-$ events  
tend to be soft and collinear, owning to the dominant infrared dynamics
of the gluon splitting (see the dashed lines in the top plots of 
Figs.~\ref{fig_res_350} and \ref{fig_res_500} below). 
In other terms, whereas to increase the value of $y_{\mathrm{cut}}$
would affect both signals only slightly, this is no longer true for the 
background.
In fact, in one increases the resolution, e.g., by a factor of five, 
to $y_{\mathrm{cut}}=0.005$,
the latter decreases by a factor of about 90(25) at 
$E_{\rm{cm}}=350(500)$ GeV. In correspondence, the typical loss for each of
 the former is less 
than a factor of 5 at both energies. 

\begin{figure}[p]
~\epsfig{file=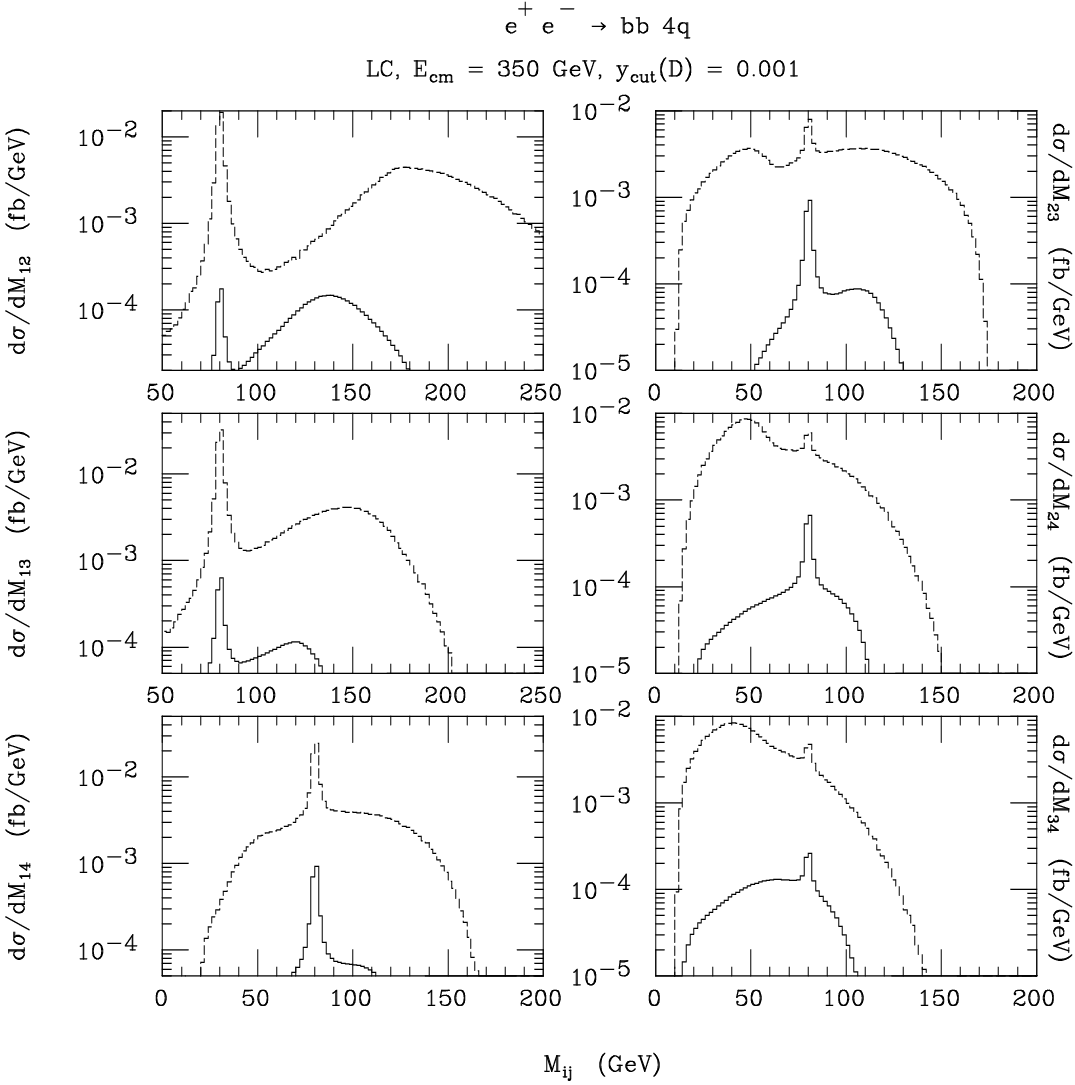,height=16.5cm,angle=0}
\caption{Differential distributions in the invariant mass of
the energy-ordered di-jet pairs, $M_{ij}$ with $i<j=1, ... 4$, for 
$e^+e^-\ar H Z$ $\ar$ $W^+W^- Z$ $\ar$
$q\bar q$$q'\bar q'$$b\bar b$ (solid lines) and 
$e^+e^-\ar W^+W^-\ar q\bar q q'\bar q' b\bar b $ (dashed lines) events,
in the MSSM for $M_A=220$ GeV and $\tan\beta=3.0$, 
with $\ycut=0.001$ in the Durham jet-finder, at $\Ecm=350$ GeV. 
The summation over all possible combinations of flavours $q$ and $q'$ 
has been performed. Spectra are normalised to the total cross sections. }
\label{fig_MW_350}
\end{figure}

Therefore, even in those cases where the background is apparently 
well above the signal,
e.g, for $M_A=220$ GeV and $\tan\beta=3.0$ GeV (corresponding to $
M_H\approx230$ GeV),
the latter being dominated at 350 GeV by $HZ\ar W^+W^-Z$ production and 
decay (the asterisk in
the upper frame of Fig.~\ref{fig_higgs}), a judicious choice of 
$y_{\mathrm{cut}}$
combined with a dedicate selection in $M_{ij}$ (with $i<j=1, ... 4$, the six 
di-jet invariant masses that can be reconstructed from the light quark jets
ordered in energy, see Fig.~\ref{fig_MW_350})
and $M_{b\bar b}$ 
(that of the two $b$ jets, see top frame of Fig.~\ref{fig_res_350}) 
around the $W^\pm$ and $Z$ masses, respectively, might allow for
the remotion of $e^+e^-\ar W^+W^-\ar q\bar q q'\bar q' b\bar b$ 
events in the spectrum of the four-light-quark invariant mass, 
in which the Higgs peak generated in the production
and decay sequence $e^+e^-\ar H Z\ar W^+W^- Z\ar q\bar q q'\bar q' b\bar b$ 
should appear
(bottom frame of Fig.~\ref{fig_res_350}). (Notice that bins are there
two GeV wide, being the expected $M_{4q}$ resolution more realistically, say, 
five times as large, so that the signal would actually be well below the 
background in this distribution, after jet-selection cuts only.) 
However, even for  
$\int{\cal L}dt=500$ fb$^{-1}$, the event rate is rather poor in this
case, about four events per year.

\begin{figure}[p]
~\epsfig{file=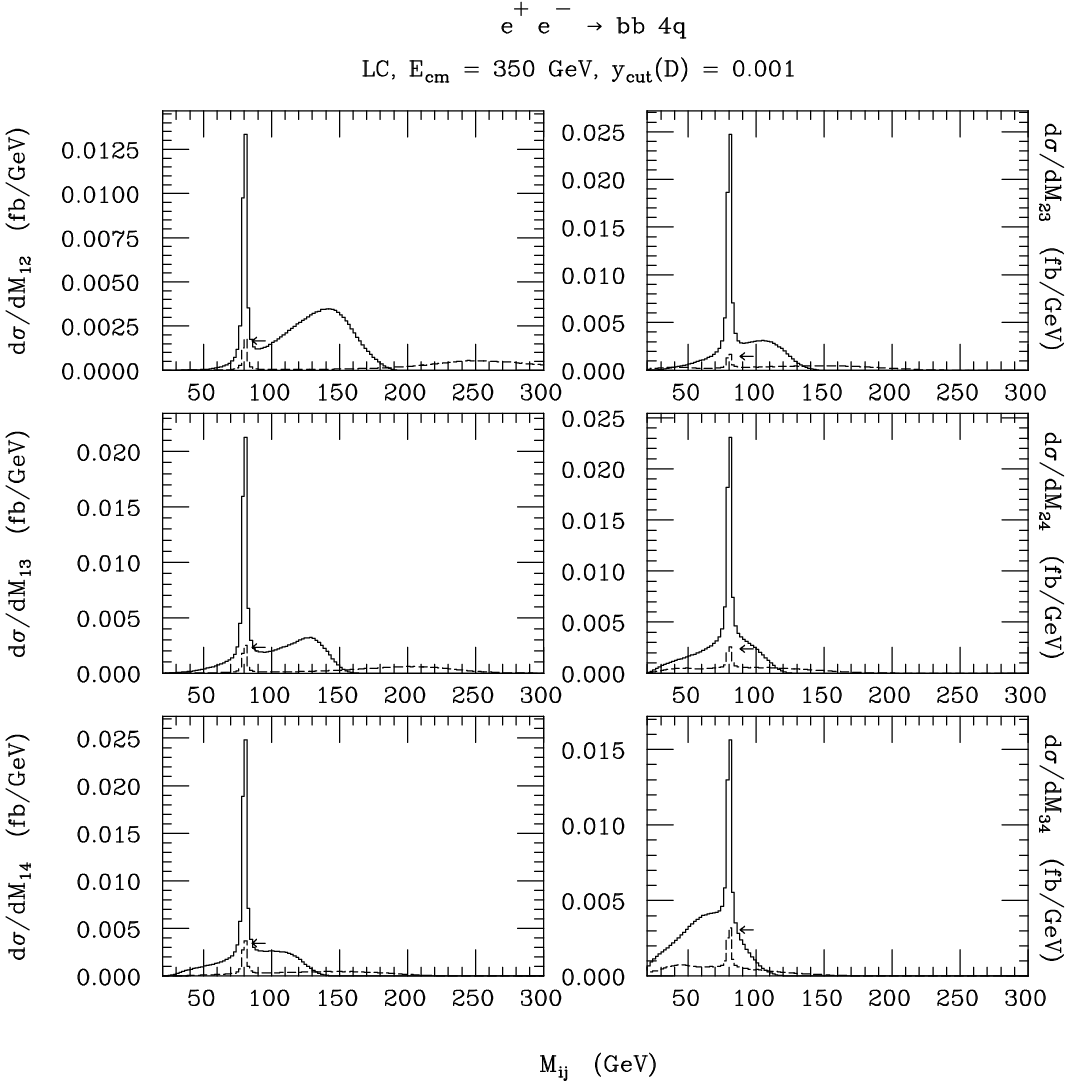,height=16.5cm,angle=0}
\caption{Differential distributions in the invariant mass of
the energy-ordered di-jet pairs, $M_{ij}$ with $i<j=1, ... 4$, for 
$e^+e^-\ar H A$ $\ar$ $W^+W^-A$ $\ar$
$q\bar q$$q'\bar q'$$b\bar b$ (solid lines) and 
$e^+e^-\ar W^+W^-\ar q\bar q q'\bar q' b\bar b $ (dashed lines,
where the peak height is denoted by an arrow) events,
in the MSSM for $M_A=220$ GeV and $\tan\beta=3.0$, 
with $\ycut=0.001$ in the Durham jet-finder, at $\Ecm=500$ GeV. 
The summation over all possible combinations of flavours $q$ and $q'$ 
has been performed. Spectra are normalised to the total cross sections. }
\label{fig_MW_500}
\end{figure}

In contrast, for other settings of the MSSM parameter spaces, the signal would
clearly be visible above the QCD noise considered here, even before the 
implementation
of the Higgs selection cuts. As illustrative example, we consider again
the same point, $M_A=220$ GeV and $\tan\beta=3.0$ GeV ($M_H\approx230$ GeV), 
but at 500 GeV, where the dominant Higgs channel now involves $HA\ar W^+W^-A$ 
production and decay
(see the asterisk in the bottom plot of 
Fig.~\ref{fig_higgs}). For such a choice,
the $M_W$ resonance is already much higher for the signal than for the 
background
(see Fig.~\ref{fig_MW_500}, where the arrows denote the height of the peaks 
for the
latter) and  both the $A$ and $H$ Breit-Wigner shapes clearly stick out
in the $M_{b\bar b}$ and $M_{4q}$ spectra, respectively 
(see Fig.~\ref{fig_res_500}),
even if the four-jet mass resolution is much larger than 2 GeV.  
In this case, the yearly production rate of the signal would be 140 events
(again, assuming  $\int{\cal L}dt=500$ fb$^{-1}$).

The discussion for the case of the light scalar Higgs of the MSSM is rather 
similar,
so we do not repeat it here. In this case, in general, the MSSM parameter space
accessible via the Higgs signature $h\ar W^+W^-\ar$  4 jets is much reduced 
though,
as only a light scalar with mass at the very upper hand of its allowed range
can decay in such a channel 
(in which one of the two $W^\pm$ bosons is off-shell):
see Fig.~\ref{fig_higgs}.

\section{Conclusions}
\label{sec_summary}

In the end, ${\cal O}(\alpha_s^2)$ decays into six-fermions 
of $W^+W^-$ pairs produced in $e^+e^-$ scatterings can be detected at
 LEP2 and a LC as well, the latter 
with $350~{\mathrm{GeV}}~\Ord\Ecm\Ord$ 500 GeV.
At both colliders though, they have little relevance in $M_W$
measurements. As for New Physics analyses, such events can be important
in a LC environment.
 Firstly, they  can produce very light gluinos at 
statistically significant rate. 
These sparticles have survived the LEP constraints and in some SUSY 
scenarios could well be the next-to-lightest ones, all other being much 
heavier. The new SUSY signals could be searched for
in six-jet samples in which only two of the jets 
reconstruct the $W^\pm$ mass. Using the
remaining four-jet subset one could either 
fit the QCD colour factors to the shape of some typical angular
distributions or exploit the tagging of a displaced vertex. 
 Secondly, they can represent an overwhelming background 
in the
search for Higgs bosons, both in the SM and in the MSSM, produced via
$e^+e^-\ar {\mathrm{Higgs}}~Z\ar W^+W^-Z$ and $e^+e^-\ar \Phi A\ar W^+W^-A$, 
where 
${\mathrm{Higgs}}=\phi,\Phi$
and $\Phi=H,h$ are the scalar particles and
 in which $W^+W^-\ar$ 4 (light) jets. However, to cut around the $Z$ 
and/or $A$ masses in the $b\bar b$ subsystem,
should in general allow one to reduce drastically such a
QCD noise in the Higgs candidate sample of six-jets with two displaced
vertices, provided a high $b$ tagging efficiency can be achieved.

The calculation of the exact matrix element for the $2\ar6$ process 
$e^+e^-\ar W^+W^-\ar q\bar q q'\bar q' q''\bar q''$ through the order 
${\cal O}(\alpha_{em}^4\alpha_s^2)$ has been performed  by means of helicity
amplitude methods. This has led to compact analytic expressions of the MEs
and to their fast implementation in a {\tt FORTRAN} code which can 
profitably be exploited in high statistic MC simulations. 

\section*{Acknowledgements}

SM is grateful to the UK PPARC for support and to
the Theoretical Physics Department of the University of Granada 
for kind hospitality while part of this work was carried out.
Many useful discussions with Ramon Mu\~noz-Tapia
during the earlier stages of this work are kindly acknowledged.

\vfill

\clearpage\thispagestyle{empty}

\clearpage\thispagestyle{empty}

\clearpage\thispagestyle{empty}

\clearpage\thispagestyle{empty}

\clearpage\thispagestyle{empty}


\clearpage\thispagestyle{empty}
\begin{figure}[p]
~\epsfig{file=rmt_higgs_350.ps,height=14cm,angle=90}
\vskip0.0001mm
~\epsfig{file=rmt_higgs_500.ps,height=14cm,angle=90}
\caption{Cross section for $e^+e^-\ar \Phi Z$ $\ar$ $W^+W^-Z$ $\ar$ 
$q\bar q$$q'\bar q'$$b\bar b$, $e^+e^-\ar \Phi A$ $\ar$ $W^+W^-A$ $\ar$
$q\bar q$$q'\bar q'$$b\bar b$ (solid, dashed, dotted and dot-dashed
lines) and  $e^+e^-\ar W^+W^-\ar q\bar q
q'\bar q' b\bar b $ (arrows) events as a function of the 
scalar Higgs masses $M_h$ and $M_H$, with
$\ycut=0.001$ in the Durham jet-finder, at $\Ecm=350$ 
(upper plot) and 500 (lower plot) GeV.
The summation over all possible combinations of flavours $q$ and $q'$
has been performed.}
\label{fig_higgs}
\end{figure}

\clearpage\thispagestyle{empty}

\clearpage\thispagestyle{empty}

\clearpage\thispagestyle{empty}
\begin{figure}[p]
~\epsfig{file=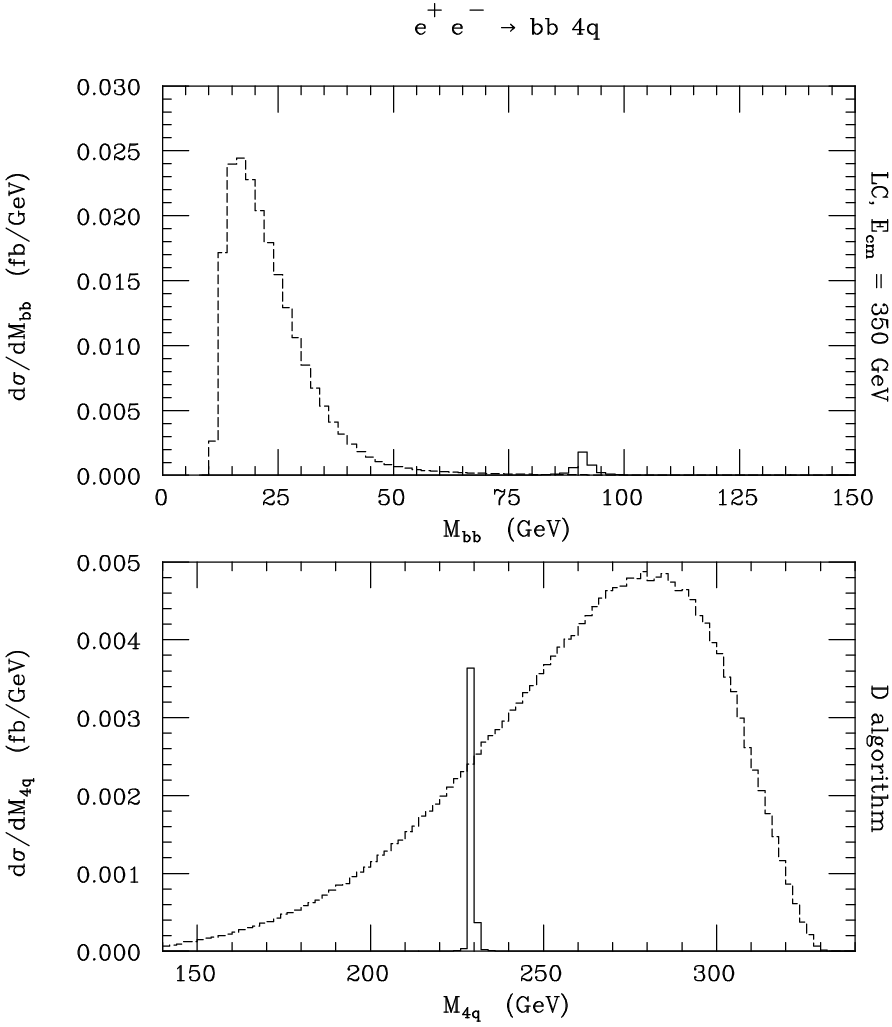,height=17cm,angle=0}
\caption{Differential distributions in the invariant mass of
the $b\bar b$, $M_{b\bar b}$ (upper plot), and 
four-light-quark, $M_{4q}$ (lower plot), systems, for
$e^+e^-\ar H Z$ $\ar$ $W^+W^- Z$ $\ar$
$q\bar q$$q'\bar q'$$b\bar b$ (solid lines) and 
$e^+e^-\ar W^+W^-\ar q\bar q q'\bar q' b\bar b $ (dashed lines) events,
in the MSSM for $M_A=220$ GeV and $\tan\beta=3.0$, 
with $\ycut=0.001$ in the Durham jet-finder, at $\Ecm=350$ GeV. 
The summation over all possible combinations of flavours $q$ and $q'$ 
has been performed. Spectra are normalised to the total cross sections. }
\label{fig_res_350}
\end{figure}

\clearpage\thispagestyle{empty}

\clearpage\thispagestyle{empty}
\begin{figure}[p]
~\epsfig{file=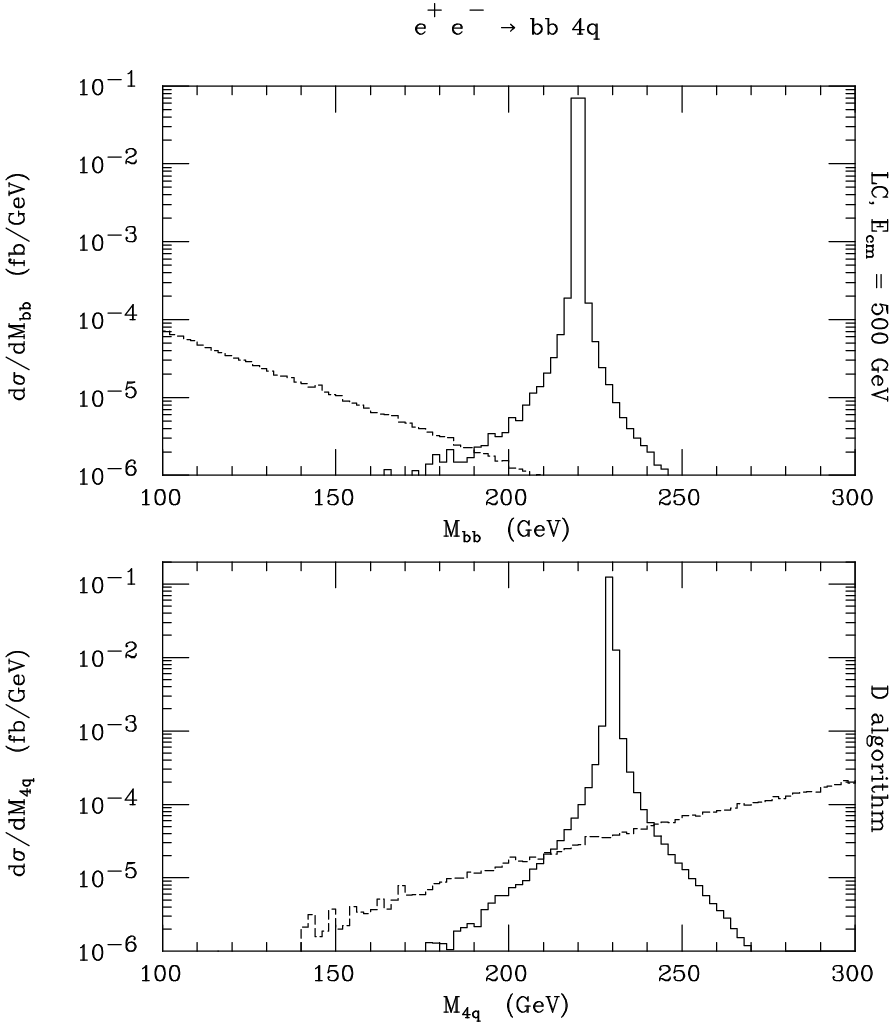,height=17cm,angle=0}
\caption{Differential distributions in the invariant mass of
the $b\bar b$, $M_{b\bar b}$ (upper plot), and 
four-light-quark, $M_{4q}$ (lower plot), systems, for
$e^+e^-\ar H A$ $\ar$ $W^+W^- A$ $\ar$
$q\bar q$$q'\bar q'$$b\bar b$ (solid lines) and 
$e^+e^-\ar W^+W^-\ar q\bar q q'\bar q' b\bar b $ (dashed lines) events,
in the MSSM for $M_A=220$ GeV and $\tan\beta=3.0$, 
with $\ycut=0.001$ in the Durham jet-finder, at $\Ecm=500$ GeV. 
The summation over all possible combinations of flavours $q$ and $q'$ 
has been performed. Spectra are normalised to the total cross sections. }
\label{fig_res_500}
\end{figure}

\end{document}